\documentclass[aps,preprint,prc,letterpaper,english,superscriptaddress,showpacs]{revtex4-1}

\usepackage[ansinew]{inputenc}
\usepackage[T1]{fontenc}
\usepackage[english]{babel}
\usepackage[usenames,dvipsnames]{color}
\usepackage{graphicx}
\usepackage{amsmath}
\usepackage{amssymb}
\usepackage{multirow}
\usepackage{dcolumn}
\usepackage[all]{xy}

\newcommand{\code}[1]{\texttt{#1}}
\newcommand{\incl}{\code{INCL4.5}}
\newcommand{\inclbare}{\code{INCL}}
\newcommand{\isabel}{\code{Isabel}}

\newcommand{\geminipp}{\code{GEMINI++}}
\newcommand{\abla}{\code{ABLA07}}
\newcommand{\gem}{\code{GEM}}
\newcommand{\smm}{\code{SMM}}

\newcommand{\etal}{et al.}
\newcommand{\proton}{\textit{p}}
\newcommand{\iron}{$^\text{56}$Fe}
\newcommand{\xenon}{$^\text{136}$Xe}

\newcommand{\geant}{\code{Geant4}}

\begin{document}

\title{The elusiveness of multifragmentation footprints in 1-GeV proton-nucleus reactions}

\author{Davide Mancusi} \email[Corresponding author. Present address: CEA, Centre de Saclay, IRFU/Service de Physique
Nucléaire, F-91191 Gif-sur-Yvette, France. E-mail address: ]{davide.mancusi@cea.fr}
\affiliation{Fundamental Interactions in Physics and Astrophysics, University of Li\`{e}ge, all\'{e}e du 6 ao\^{u}t 17,
  b\^{a}t.~B5, B-4000 Li\`{e}ge 1, Belgium}
\author{Alain Boudard}
\affiliation{CEA, Centre de Saclay, IRFU/Service de Physique Nucléaire, F-91191 Gif-sur-Yvette, France}
\author{Joseph Cugnon}
\affiliation{Fundamental Interactions in Physics and Astrophysics, University of Li\`{e}ge, all\'{e}e du 6 ao\^{u}t 17,
  b\^{a}t.~B5, B-4000 Li\`{e}ge 1, Belgium}
\author{{Jean-Christophe} David}
\affiliation{CEA, Centre de Saclay, IRFU/Service de Physique Nucléaire, F-91191 Gif-sur-Yvette, France}
\author{Thomas Gorbinet}
\affiliation{CEA, Centre de Saclay, IRFU/Service de Physique Nucléaire, F-91191 Gif-sur-Yvette, France}
\author{Sylvie Leray}
\affiliation{CEA, Centre de Saclay, IRFU/Service de Physique Nucléaire, F-91191 Gif-sur-Yvette, France}

\date{\today}

\begin{abstract}
  We use the tools of hybrid intranuclear-cascade/nuclear-de-excitation models to evaluate the sensitivity of several
  physical observables to the inclusion of a multifragmentation stage in the de-excitation chain and assess the need for a
  multifragmentation model in the quantitative description of \proton+\iron\ and \proton+\xenon\ reactions at 1-GeV
  incident energy. We seek clear signatures of multifragmentation by comparing different state-of-the-art de-excitation
  models coupled with intranuclear-cascade models and by focusing on discriminating observables such as correlations and
  fragment longitudinal-velocity distributions. None of the considered observables can be unambiguously interpreted as a
  multifragmentation footprint. The experimental data are best described as originating from sequential binary
  decays. However, no de-excitation model can reproduce the experimental longitudinal-velocity distributions from 1-GeV
  \proton+\xenon.
\end{abstract}


\maketitle

\section{Introduction}

Multifragmentation is generally considered as the quasi-simultaneous break-up of highly-excited nuclear matter into
clusters and unbound nucleons. Interest towards this phenomenon was first triggered by anomalously large production cross
sections of intermediate-mass fragments (abbreviated as IMFs and defined as $3\leq Z\leq 10$ for the purpose of this
paper) from intermediate-energy heavy-ion collisions (see Ref.~\onlinecite{epj30_1} for a collection of recent reviews).
The earliest theoretical explanations suggested to interpret the typical power-law distribution of fragment masses as a
signature of liquid-vapour equilibrium of nuclear matter near the critical temperature. In this scheme, nuclear clusters
are simultaneously formed, with large multiplicities, as condensation droplets of a vapour of nucleons. However, as the
initial enthusiasm over liquid-vapour multifragmentation faded and other candidate models (e.g.\ statistical
multifragmentation, spinodal instability and even sequential binary decays) were put forward to explain the data, it was
quickly realised that the power-law signature was by no means unique of liquid-vapour multifragmentation. Remarkably, even
simple percolation models are able to reproduce most of the features of the observed IMF
distributions. 
Therefore,
other observables must be sought if one wishes to discriminate among the proposed IMF production mechanisms, which are
anyway not necessarily mutually exclusive.

One of the main difficulties of multifragmentation studies based on heavy-ion reactions is that there is considerable
theoretical uncertainty on the reaction dynamics and on the importance of collective effects such as deformation or
compression. In nucleon-induced reactions, on the other hand, it is difficult to imagine that the collective state of the
system can be strongly perturbed. Since it had been known for a long time that IMFs could also be produced in
nucleon-induced reactions, multifragmentation studies were also performed on these better-understood systems, although the
excitation energies that can be reached by this method are typically lower.

Today, the importance of multifragmentation in nucleon-induced reactions is the subject of a long-standing
discussion. While it is generally accepted that multifragmentation will eventually set in at high projectile energy, due
to the increasing energy transfer from the projectile to the target nucleus, it is not yet clear whether and to what
extent multifragmentation needs to be postulated for a reliable quantitative description of reactions around 1~GeV, a
region which is most interesting for technical applications such as Accelerator-Driven Systems (ADS)
\cite{ait_abderrahim-myrrha}, radioprotection in space \cite{durante-radiation} and shielding at accelerators
\cite{satif10}.

The recent IAEA-promoted ``Benchmark of Spallation Models'', which focused on the 60--3000-MeV incident-energy range,
represents an effort
\textit{``to assess the prediction capabilities of the spallation models used or that could be used in the future in
  high-energy transport codes; to understand the reason for the success or deficiency of the models in the different mass
  and energy regions or for the different exit channels; to reach a consensus, if possible, on some of the physics
  ingredients that should be used in the models''}
\cite{david-intercomparison,*leray-intercomparison,*intercomparison-website}.
The benchmark saw the participation of seventeen spallation models, all of which were
couplings of a dynamical reaction model (intranuclear cascade, quantum molecular dynamics\ldots) and a statistical-decay
model, with the possible presence of an intermediate pre-equilibrium stage. Since not all the participating models include
a multifragmentation stage, it is in principle possible to study the benchmark results and estimate the sensitivity of the
benchmark endpoints (isotopic production cross sections, excitation curves, neutron-multiplicity distributions,
double-differential cross sections for neutrons, light charged particles and pions) to the multifragmentation process. In
particular, by comparing the predictions of different de-excitation models coupled with a fixed dynamical stage, one can
extract precious information about the influence of de-excitation alone.

However, previous studies have already indicated that inclusive observables, such as double-differential nucleon spectra
or nuclide yields, are rather insensitive to the inclusion of a multifragmentation stage in the de-excitation chain
\cite{legentil-fe}. Hence, characteristic signatures of multifragmentation must be sought among other, more discriminating
observables. The impact of a multifragmentation stage in the de-excitation chain can in principle be assessed by comparing
calculation results with experimental data.

The goal of the present work is to identify possible signatures of multifragmentation by studying nucleon-induced
reactions with the tools of coupled intranuclear-cascade/de-excitation models. We shall focus on the 1-GeV \proton+\iron\
and \proton+\xenon\ reactions, which have been the object of recent studies
\cite{villagrasa-fe,legentil-fe,napolitani-xe_xsec,napolitani-fe,napolitani-xe_velocity}. 
The small mass of the \proton+\iron\ system leads to the production of a limited number of nuclides. Several de-excitation
mechanisms can contribute to a given nuclide yield, making it more difficult to extract an unambiguous multifragmentation
signature from a background of de-excitation residues and/or direct IMF emissions.  On the other hand, the
multifragmentation threshold may be more easily attained in \proton+\iron, which realises higher excitation energies per
nucleon. Thus, the two systems studied are complementary. Heavier systems are excluded from the present study, in order to
avoid the conceptual and technical complications connected to the competition between light-fragment emission and fission.

The paper is organised as follows: in Sec.~\ref{sec:model-overview} we give a brief overview of the models used for the
study. Sec.~\ref{sec:sensitivity-cascade} examines the sensitivity of the considered observables to the choice of the
intranuclear-cascade model. Sec.~\ref{sec:incl-cross-sect} discusses the available inclusive residue-production data and
how they are reproduced by the different de-excitation models.  Sec.~\ref{sec:spal-corr} presents the model predictions
for the SPALADIN correlation data-set for \proton+\iron\ \cite{legentil-fe}. Sec.~\ref{sec:long-veloc-distr} discusses
longitudinal-velocity distributions measured in inverse kinematics at GSI, Darmstadt, Germany
\cite{napolitani-fe,napolitani-xe_velocity}. Sec.~\ref{sec:time-interv-betw} discusses time intervals between fragment
emissions in \incl/\geminipp. Finally, Sec.~\ref{sec:conclusions} summarises our conclusions.

\section{Model overview}\label{sec:model-overview}

All the calculations presented in this paper were performed using a coupled intranuclear-cascade/statistical-de-excitation
model. We used two intranuclear-cascade model (\incl\ and \isabel) coupled with three different de-excitation models
(\abla, \geminipp\ and \smm). Since the focus of this paper is on de-excitation, we will limit ourselves to directing the
reader to the relevant publications for details about the physics of the cascade models.

\subsection{Cascade models}\label{sec:cascade-models}

The Liège Intranuclear Cascade model (\inclbare) \cite{boudard-incl,cugnon-incl45_nd2010} is one of the
most refined existing tools for the description of nucleon-, pion- and light-ion-induced reactions in the 150--3000-MeV
incident energy range. The model is currently maintained and developed jointly by the University of Liège (Liège, Belgium)
and CEA (Saclay, France). It can describe the emission of nucleons and pions; light clusters (up to $Z=5$, $A=8$ with the
default program options) can also be produced through a dynamical phase-space coalescence algorithm.
The \inclbare\ model is not to be considered as adjustable. It does contain parameters, but they are
either taken from known phenomenology (such as the matter density radius of the nuclei) or have been adjusted once for all
(such as the parameters of the Pauli blocking or those that determine the coalescence module for the production of the
light charged clusters). The predictions of \inclbare\ concerning those observables that can be confronted directly to
experiment, namely the high energy parts of particle spectra, are of rather good quality, as it was recently shown
\cite{cugnon-incl45_nd2010}. The \inclbare/\abla, \inclbare/\geminipp\ and \inclbare/\smm\ combinations were also recognised among the
best-performing participants of the IAEA ``Benchmark of Spallation Models''
\cite{david-intercomparison,*leray-intercomparison,*intercomparison-website}. The present work
is based on the \inclbare\ version that was used for the IAEA benchmark, plus some minor bug fixes; this
version is known as \incl.

An older version of the \inclbare\ model, known as \inclbare\code{4.2}, was employed for studying the SPALADIN correlation
data-set \cite{legentil-fe}. The most important differences between \inclbare\code{4.2} and \code{4.5} are reviewed in
Ref.~\onlinecite{cugnon-incl45_nd2010} and include the introduction of the cluster-coalescence algorithm, energy- and
isospin-dependent potentials for nucleons and pion potentials, as well as an improvement of Pauli blocking. More details
are given in the reference above.

The \isabel\ model \cite{yariv-isabel1,*yariv-isabel2}, no longer developed, has vastly contributed to the understanding of
nucleon-nucleus and nucleus-nucleus reactions. It is mainly used in the present work to highlight the sensitivity of the
studied observables to cascade.


\subsection{De-excitation models}\label{sec:de-excitation-models}

\subsubsection{\abla}

The \abla\ model \cite{kelic-abla07} is maintained and developed by the CHARMS group at GSI, Darmstadt, Germany. The model
contains a multifragmentation sub-module, which is triggered only if the temperature of the compound nucleus exceeds a
mass-dependent (as suggested by Natowitz \etal\ \cite{natowitz-systematics}) freeze-out threshold:
\begin{equation}
  T_\text{freeze-out}(A)=\max\left(5.5,\,9.337\cdot\exp(-2.82\cdot10^{-3}A)\right)\text{ MeV.}\label{eq:freeze-out}
\end{equation}
In that case, the system breaks up into fragments whose mass is distributed according to an empirical power-law spectrum,
and whose momenta carry Goldhaber-type and thermal contributions. Coulomb repulsion among multifragmentation products is
accounted for in a simplified manner. The excitation energies of the resulting fragments are determined by assuming
thermal equilibrium at the freeze-out temperature. Subsequent de-excitation of the multifragmentation products is assumed
to be purely binary. If the multifragmentation module is not triggered, the initial compound nucleus directly enters the
secondary de-excitation phase.

During secondary de-excitation, emission of any stable nucleus up to half the mass of the compound nucleus is possible,
and it is quantitatively described by the Weisskopf-Ewing evaporation formalism \cite{weisskopf-evaporation}. Above the
Businaro-Gallone point, competition with fission is treated dynamically and it is based on solutions of the Fokker-Planck
equation for collective deformation of the nucleus over the fission barrier. \abla's fission module is among the most
sophisticated models available on the market, but it is only of marginal interest for the systems studied in this paper.

Finally, subsequent binary decays are assumed to be independent; in particular, Coulomb interactions among particles
produced in different decays are neglected. This assumption is customary in binary de-excitation models.

A less-sophisticated version of the \code{ABLA} model was considered for the study of the SPALADIN correlations
\cite{legentil-fe} and was found unable to reproduce the measured residue-production cross sections in \proton+\iron. It
did not include multifragmentation, nor evaporation of fragments heavier than alpha particles. More details about the
differences between the two versions can be found in Ref.~\onlinecite{kelic-abla07}.

\subsubsection{\geminipp}\label{sec:geminipp}

The \geminipp\ model, developed by R.~J.~Charity \cite{charity-gemini++}, represents an effort to describe nuclear
de-excitation uniquely in terms of binary decays. No simultaneous break-up is allowed. Multi-fragment events can of course
be produced by sequences of binary fragment emissions; as in the case of \abla, Coulomb interactions among particles
emitted in different decays are neglected. Emission of light particles ($Z\leq3$ by default) is described by the
Hauser-Feshbach evaporation formalism \cite{hauser-evaporation}; Moretto's conditional-saddle-point formalism
\cite{moretto-binarydecay} with Sierk's finite-range barriers \cite{sierk-asyBar} is used for complex-fragment
emission. For heavy systems, the fission width is calculated using a refined Bohr-Wheeler approach
\cite{mancusi-gemini++_fission}.

\geminipp's asymmetric-fission module has recently been improved \cite{mancusi-fusion11} to describe fragment yields from
fusion and spallation reactions with the same parameter set. To this end, it was necessary to augment Sierk's barriers by
a constant shift of 7~MeV, which can be interpreted as the difference in Wigner energy between the mother nucleus and the
nascent fragments. However, this interpretation is not devoid of complications; see the relevant papers for more details.

\subsubsection{\smm}

The Statistical Multifragmentation Model (\smm) \cite{bondorf-multifragmentation,botvina-smm}, presently maintained by
A.~S.~Botvina, is one of the most successful and widely applied multifragmentation models. Like in \abla\, the first
possible decay stage is the simultaneous break-up of the thermalised source in a number of hot fragments and
particles. Unlike \abla's semi-empirical approach, \smm\ always enters this sub-module and samples break-up configurations
according to their thermodynamical weight in a given freeze-out volume (taken to be three times the normal nuclear
volume). Thus, at low excitation energy, single-fragment configurations (i.e.\ compound nucleus) naturally dominate the
multifragmentation phase; the importance of multi-fragment configurations smoothly increases and starts dominating the
thermodynamical weight around 3~$A$MeV excitation energy. Mass, charge, excitation energy and momentum of the emerging hot
fragments are sampled respecting conservation laws. Coulomb acceleration is then accounted for by solving the Hamilton
equations for the propagation of the fragments in their mutual Coulomb field.

Secondary de-excitation is then applied to the hot fragments. If they are sufficiently light ($A\leq16$), the Fermi
break-up model is applied. Otherwise, according to a modified Weisskopf-Ewing \cite{weisskopf-evaporation} scheme, they
can evaporate particles up to $^\text{18}$O. Fission is described by the Bohr-Wheeler model \cite{bohr-fission}.

Note that the \smm\ version used for the present work employs slightly different evaporation barriers compared to the IAEA
benchmark. Barriers are computed using the standard formula
\[
B=(1.44\text{ MeV}\cdot\text{fm})\cdot \frac{Z_1Z_2}{r_0(A_1^{1/3}+A_2^{1/3})}\text.
\]
In the IAEA benchmark, $r_0=1.5$~fm was used. In the present work, $r_0$ is determined as
\[
r_0=2.173\cdot\frac{1+6.103\cdot 10^{-3}Z_1Z_2}{1+9.443\cdot 10^{-3}Z_1Z_2}\text{ fm.}
\]
This difference is marginal as far as IMF cross sections are concerned.

\section{Basic cascade results}\label{sec:sensitivity-cascade}

\begin{figure}
  \centering
  \includegraphics[height=0.32\linewidth]{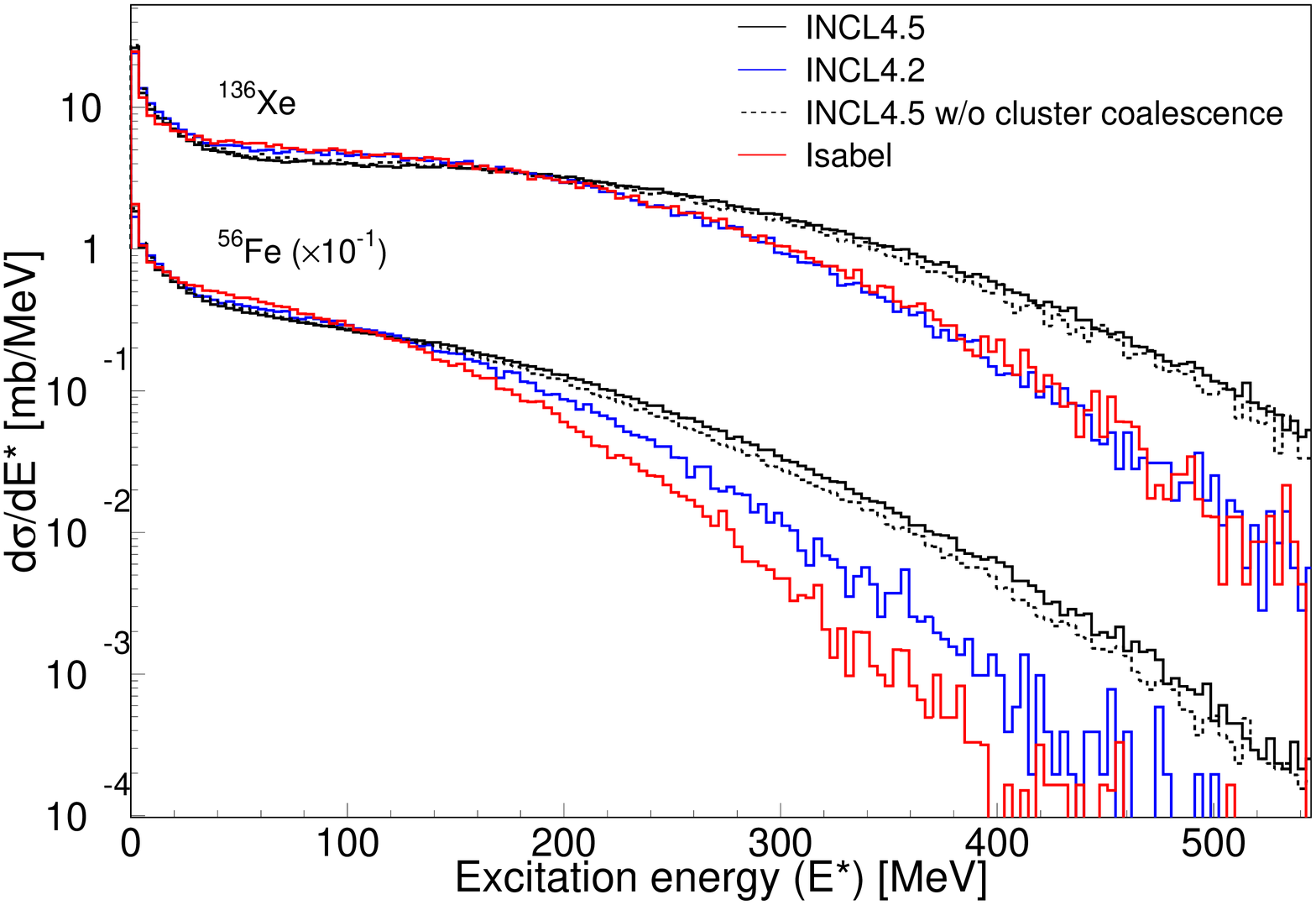}\hspace{1em}%
  \includegraphics[height=0.32\linewidth]{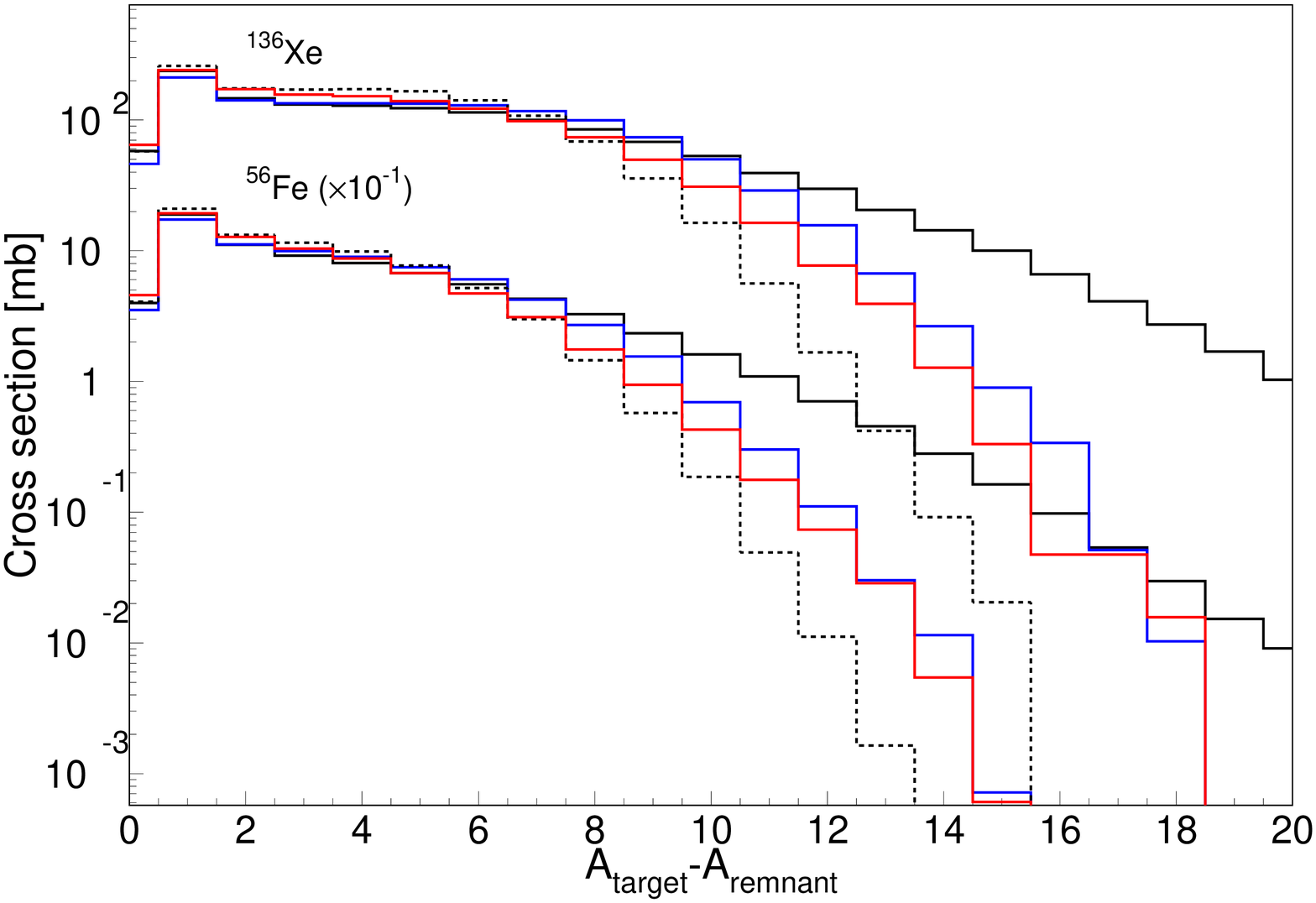}
  \caption{Comparison of excitation-energy (left panel) and mass-loss (right panel) distributions for remnants of the
    1-GeV \proton+\iron\ and \proton+\xenon\ reactions, as calculated by different cascade models.}
  \label{fig:p_remnant}
\end{figure}



\begin{table}
  \centering
  \begin{tabular}{c||c|c|c|c}
     &\multirow{2}{*}{\incl}&\multirow{2}{*}{\isabel}&\incl&\multirow{2}{*}{\inclbare\code{4.2}}\\
     &&&(no clusters)&\\
    \hline
    $\sigma^\text{reac}$ (mb)&779 & 740 &779 & 742\\
    $\langle E^*\rangle$ (MeV)& 91.5 &66.6 &85.9 & 75.5\\
    $\langle A_\text{remnant}\rangle$& 52.7 &53.4 & 53.5 & 53.0 \\
    $\langle E^*/A_\text{remnant}\rangle$ (MeV)&1.82 &1.30 &1.65 & 1.49
  \end{tabular}
  \caption{Reaction cross section, average remnant excitation energies (total and per nucleon) and average remnant mass
    predicted by \incl, \isabel, \incl\ without cluster coalescence, and \inclbare\code{4.2} for 1-GeV proton-induced
    reactions on \iron.}
  \label{tab:fe56_characteristics}
\end{table}

\begin{table}
  \centering
  \begin{tabular}{c||c|c|c|c}
     &\multirow{2}{*}{\incl}&\multirow{2}{*}{\isabel}&\incl&\multirow{2}{*}{\inclbare\code{4.2}}\\
     &&&(no clusters)&\\
    \hline
    $\sigma^\text{reac}$ (mb)& 1377 & 1332  & 1381 & 1327 \\
    $\langle E^*\rangle$ (MeV)& 139.2  &116.3 & 132.6 & 113.5 \\
    $\langle A_\text{remnant}\rangle$& 131.4  & 132.4  & 132.6  & 131.7 \\
    $\langle E^*/A_\text{remnant}\rangle$ (MeV)& 1.08 &0.89 &1.02  & 0.88
  \end{tabular}
  \caption{Same as Table~\ref{tab:fe56_characteristics}, for 1-GeV \proton+\xenon.}
  \label{tab:xe136_characteristics}
\end{table}

As a first, basic comparison, Fig.~\ref{fig:p_remnant} shows distributions of excitation energy and mass loss of cascade
remnants of 1-GeV \proton+\iron\ and \proton+\xenon. Already at this stage, it is possible to observe
that \incl\ and \isabel\ are not equivalent. \incl\ produces on average hotter and lighter remnants than
\isabel, although differences in excitation energy are smaller for \xenon\ than for \iron, as quantified in
Tables~\ref{tab:fe56_characteristics} and \ref{tab:xe136_characteristics}.  Note that the reaction cross sections
predicte bdy the two codes differ by only a few percent; thus, differences in the remnant characteristics must trace back
to different cascade histories.

Previous investigations had found that \inclbare's and \isabel's excitation-energy distributions for \proton+\iron\ were
remarkably similar \cite{legentil-fe}. The claim concerned version \code{4.2} of the \inclbare\ code \cite{boudard-incl},
which is represented in Fig.~\ref{fig:p_remnant} by the blue lines. This state of affairs was evidently modified by later
developments of the \inclbare\ code \cite{cugnon-incl45_nd2010}.  We stress that the similarity between
\inclbare\code{4.2} and \isabel's results can at least partly be explained by the very similar physics content of the
models. Excitation energies are also sensibly larger in \incl\ than in \inclbare\code{4.2}. However, this difference
cannot be simply ascribed to a single cause, but rather represents the combined effect of several new physics
ingredients, such as energy-dependent nucleon potentials and pion potentials.

One major difference between \isabel\ and \incl\ (and between \inclbare\code{4.2} and \incl) is \incl's ability to
dynamically produce light charged composite particles (see Sec.~\ref{sec:cascade-models}). Fig.~\ref{fig:p_remnant}
depicts \incl's predictions with (solid black lines) and without (dotted black lines) cluster coalescence.  One
immediately observes that \isabel's mass distributions are remarkably similar to those predicted by \incl\ without
coalescence.  Clearly \incl's cluster emission algorithm reduces the remnant mass, but does not sensibly affect the
excitation-energy distribution. This is quite well understood: escaping clusters typically extract spectator nucleons
close to the Fermi sea, thereby reducing the mass of the remnant but without significantly affecting its excitation
energy. This result is also consistent with the documented behaviour of an older, less-refined clustering algorithm
\cite{boudard-incl4.3}.

Therefore, we conclude this section by observing that, contrary to what was claimed in Ref.~\onlinecite{legentil-fe}, the
choice of the intranuclear-cascade model \emph{does} have some importance. In what follows, we shall discuss how the
differences in remnant distributions are reflected in the residue-production cross sections, and how these observables can
guide us in selecting the cascade model that should be used for the study of more discriminating observables.

\section{Residue-production cross sections}\label{sec:incl-cross-sect}
 
\begin{figure}
  \centering
  \includegraphics[height=0.35\linewidth]{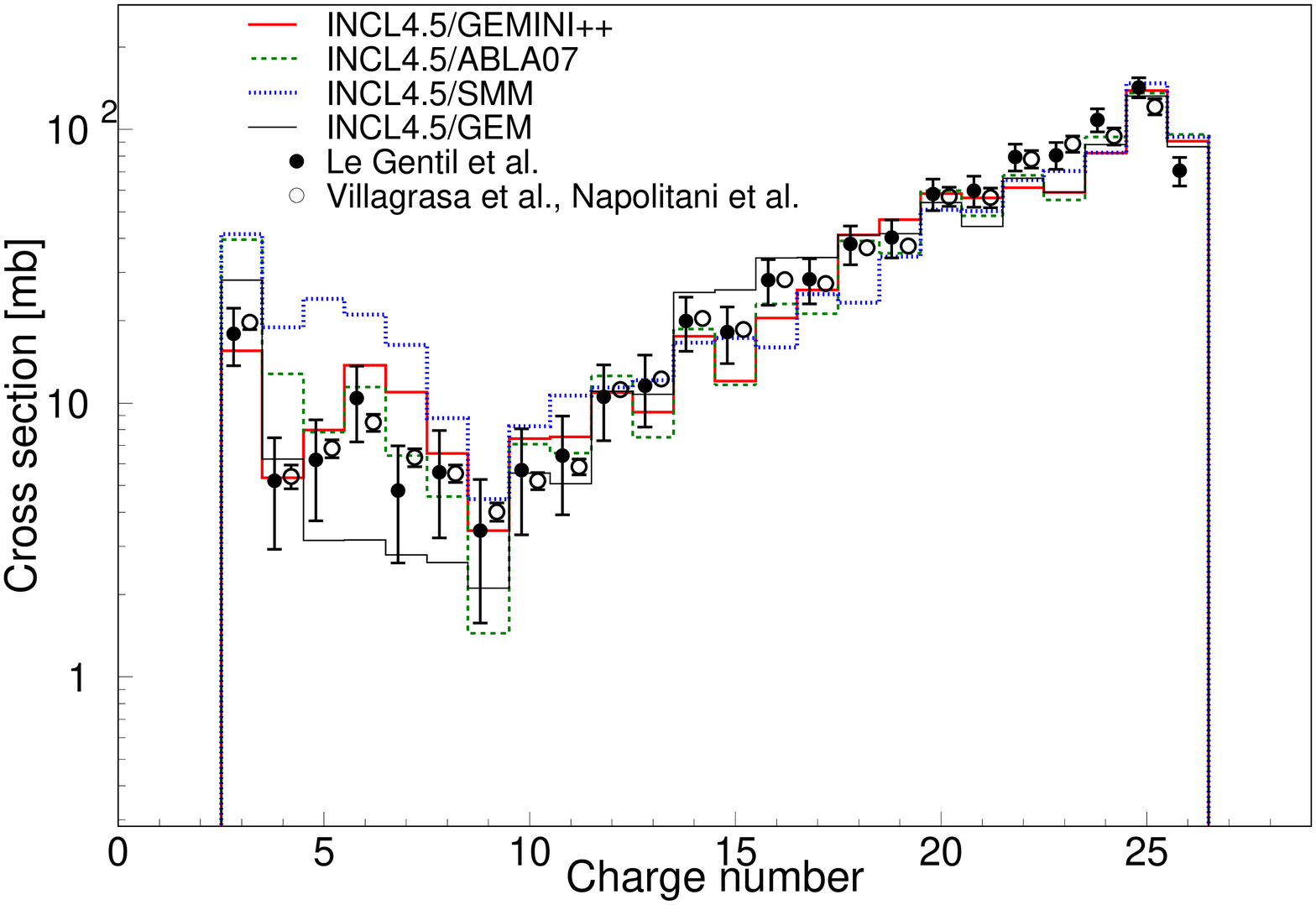}\includegraphics[height=0.35\linewidth]{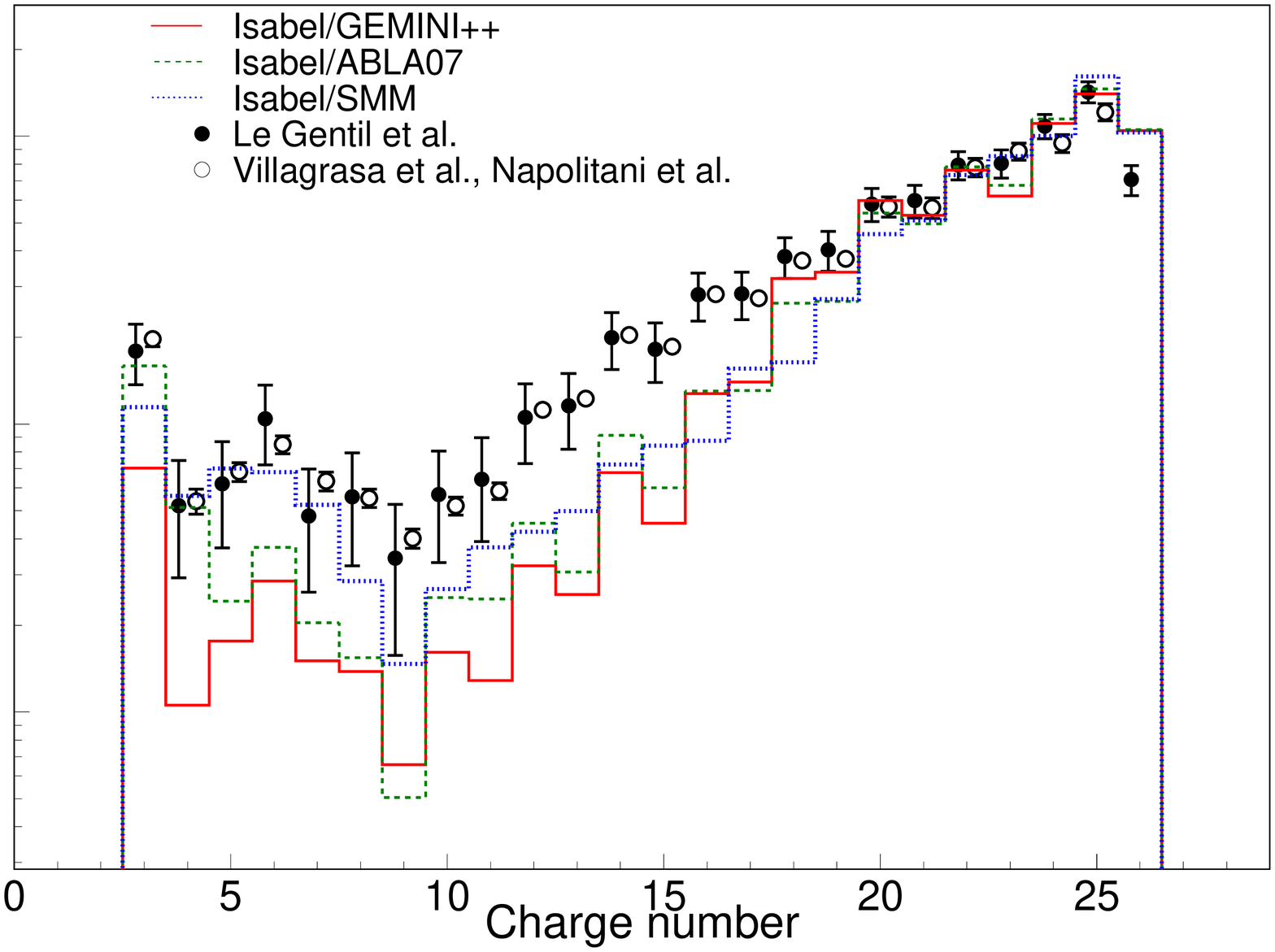}
  \caption{Inclusive residue-production cross sections for 1-GeV \proton+\iron, as a function of the nuclide charge. Left
    panel: intranuclear cascade simulated by \incl. Right panel: \isabel. Experimental data from
    Refs.~\onlinecite{legentil-fe}, \onlinecite{villagrasa-fe} and \onlinecite{napolitani-fe}.}
  \label{fig:p_fe_inclusive}
\end{figure}

\begin{figure}
  \centering
  \includegraphics[height=0.35\linewidth]{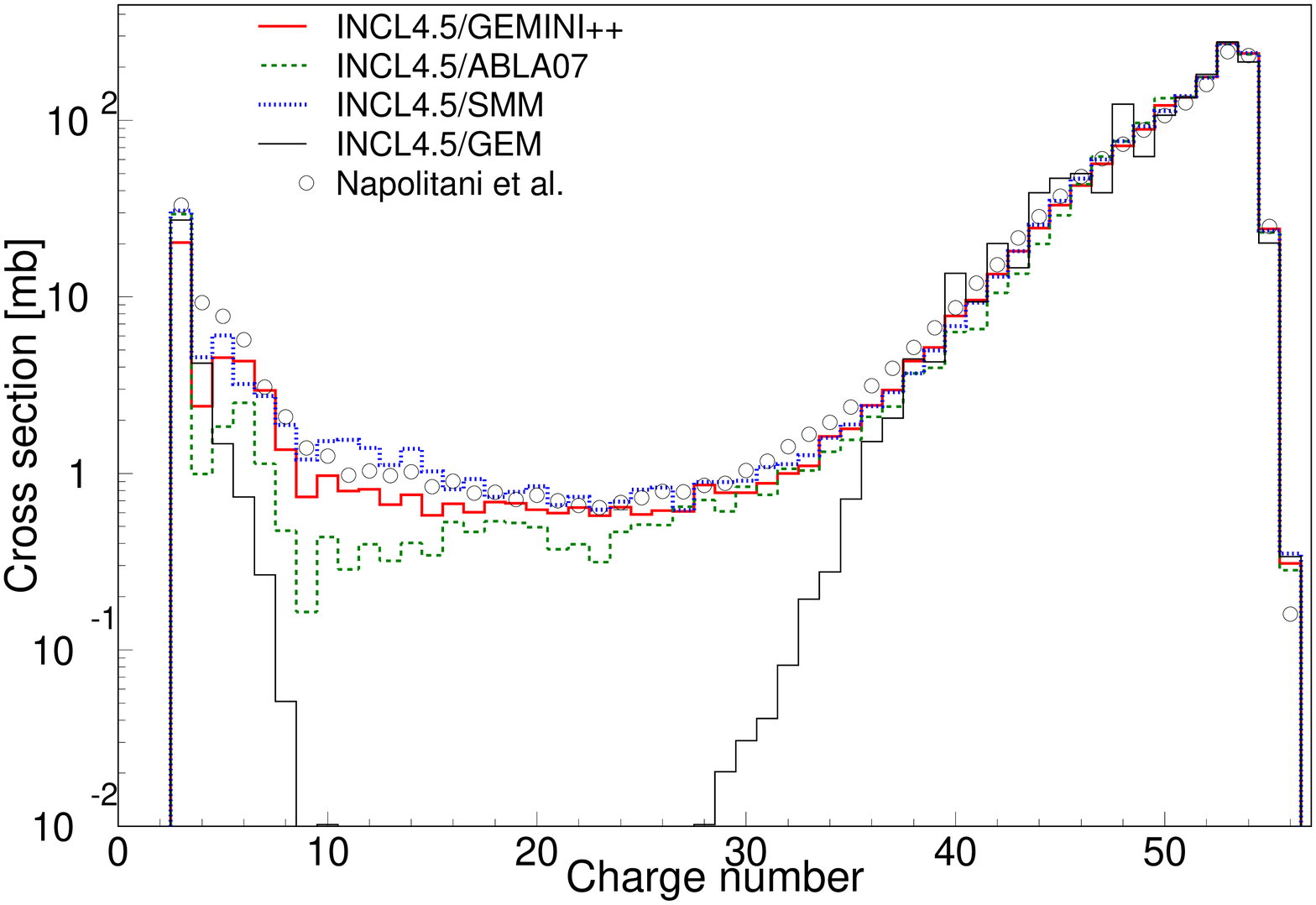}\includegraphics[height=0.35\linewidth]{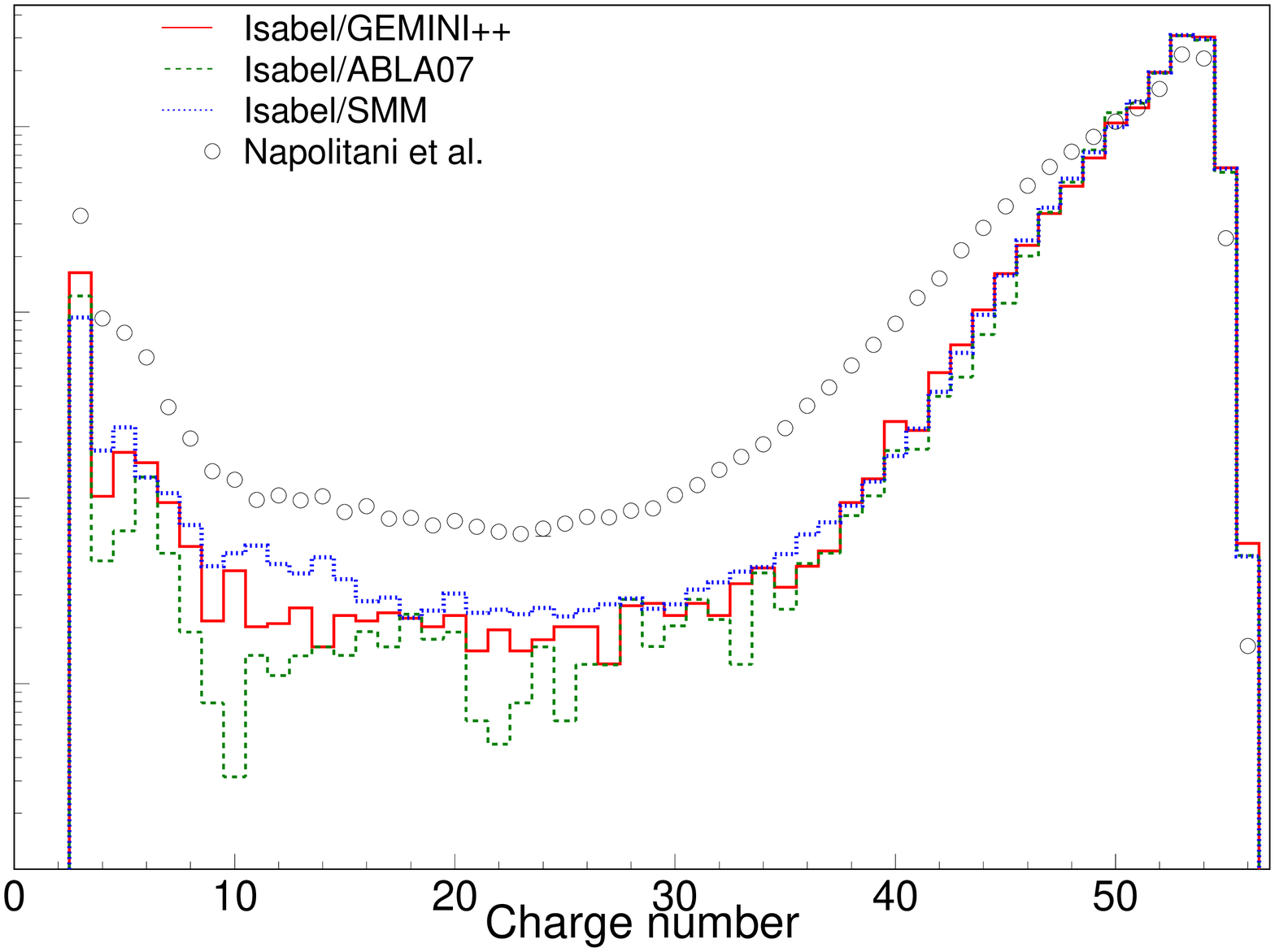}
  \caption{Same as Fig.~\ref{fig:p_fe_inclusive}, for 1-GeV \proton+\xenon. Experimental data from
    Ref.~\onlinecite{napolitani-xe_xsec}.}
  \label{fig:p_xe_inclusive}
\end{figure}

Figures~\ref{fig:p_fe_inclusive} and~\ref{fig:p_xe_inclusive} show the predicted residue-production cross sections as a
function of the nuclide charge, compared with the measurements obtained with the SPALADIN apparatus \cite{legentil-fe} or
at the FRagment Separator (FRS) \cite{villagrasa-fe,napolitani-fe,napolitani-xe_xsec}. The two experiments have different
acceptance cuts. The SPALADIN data, which will be discussed in detail in Sec.~\ref{sec:spal-corr}, present a kinematical
cut on particles with large longitudinal velocities with respect to the initial \iron\ nucleus, which are mostly nucleons
and light particles. The acceptance was estimated to be virtually complete for $Z\geq4$. In the FRS data, on the other
hand, only a selected number of isotopes were measured. For the model curves in Figs.~\ref{fig:p_fe_inclusive} and
\ref{fig:p_xe_inclusive}, we have chosen to define the residue-production cross sections as the sums of the calculated
isotopic residue-production cross sections over the nuclides observed in the FRS experiments. No kinematical cut was
applied.  Since the SPALADIN and FRS data sets are largely compatible, as it clearly appears from
Fig.~\ref{fig:p_fe_inclusive}, we do not expect this choice to bias our analysis.

For a given de-excitation stage, the curves reflect the differences in the cascade output. Note that calculated cross
sections for residues close to the target (say $Z\geq20$ for \iron\ and $Z\geq45$ for \xenon) are almost independent of
the choice of the de-excitation model and are dominated by the cascade model; in particular, they are consistently better
reproduced by \incl\ than \isabel. We consider that an accurate prediction of these cross sections is a crucial
prerequisite that the cascade model must satisfy if it is to be used for the study of more exclusive and discriminating
observables. Thus, in Sec.~\ref{sec:spal-corr} and following, we will only retain the \incl\ model for the analysis of
fragment correlations and velocity distributions.

\incl's hotter and lighter remnants also lead on average to lighter residues. Note that the IMF cross sections in
\proton+\iron\ and the $10\lesssim Z\lesssim30$ cross-section plateau in \proton+\xenon\ are better reproduced by the
\incl/de-excitation combinations, while \isabel\ consistently underestimates the \xenon\ cross sections by a factor of
about three. Comparison with the excitation-energy distributions in Fig.~\ref{fig:p_remnant} suggests that these cross
sections are associated with highly-excited remnants. Likewise, \isabel\ consistently underestimates cross sections with
$10\lesssim Z\lesssim18$ in \proton+\iron.

The sensitivity of the inclusive residue-production cross sections to cascade can be further illustrated by considering
the results obtained with the \gem\ de-excitation model \cite{furihata-gem}, coupled with \incl\ (left panes of
Fig.~\ref{fig:p_fe_inclusive} and \ref{fig:p_xe_inclusive}). In a previous study \cite{legentil-fe}, the
\inclbare\code{4.2}/\gem\ combination was excluded from the study of SPALADIN correlations because it was unable to
reproduce the IMF-production cross sections in \proton+\iron. New \incl/\gem\ calculations predict IMF cross sections that
are about a factor of 3 higher than the \code{INCL4.2}/\gem\ and in acceptable agreement with the experimental data; this
is due to \incl's different excitation-energy and remnant-mass distributions.  However, the plateau cross sections in
\proton+\xenon\ are underestimated by at least three orders of magnitude by \incl/\gem. Thus, we also exclude the \gem\
de-excitation model from this study.

If we now focus on a fixed cascade model (e.g.\ \incl), we can observe that the three de-excitation models produce similar
charge distributions. In this sense, we confirm that residue-production cross sections are rather insensitive to the
de-excitation mechanism. However, we remark that de-excitation models present free parameters that can be adjusted to help
reproduce the residue-production cross sections. The \proton+\iron\ data-set, in particular, is a popular benchmark for
spallation models (cascade/de-excitation) due to its good accuracy. The \geminipp\ parameters connected with asymmetric
fission were admittedly fitted to the \proton+\iron\ and \proton+\xenon\ residue-production cross sections, among other
data-sets \cite{mancusi-fusion11}. Thus, Figs.~\ref{fig:p_fe_inclusive} and \ref{fig:p_xe_inclusive} can deceptively lead
to underestimate the sensitivity of residue-production cross sections to the de-excitation model.

The sensitivity to de-excitation can be further appreciated by analysing how different de-excitation mechanisms contribute
to the residue-production cross section.

\subsection{Production mechanism in the de-excitation models}\label{sec:production-mechanism}

It is instructive to study how the different de-excitation models reconstruct the residue-production cross sections as the
sum of different production mechanisms. However, we need to introduce this discussion by a few important remarks. Firstly,
different de-excitation models have different, possibly non-overlapping sets of production mechanisms
(Sec.~\ref{sec:de-excitation-models}); thus, each partition must be seen as model-dependent and cannot be directly
compared to experimental data or to other partitions. Secondly, although all models internally construct some kind of
de-excitation-history tree, only a limited, model-dependent amount of information about the decay history is readily
available to the user. Figure~\ref{fig:mechanisms} summarises how production mechanisms are partitioned in each model;
details about each partitioning will be given in the model-specific discussions that follow. Each de-excitation mechanism
is assigned a colour, which is consistently used in Figs.~\ref{fig:p_smm_mechanism}--\ref{fig:p_geminipp_mechanism},
\ref{fig:velocity_fe_components_selection} and \ref{fig:velocity_xe_components_selection}. We attempted to assign similar
colours to similar mechanisms. In some cases, a specific mechanism in one model can be considered equivalent to another
mechanism or to the sum of other mechanisms in another model. These cases are indicated by arrows and boxes in
Fig.~\ref{fig:mechanisms}.

\definecolor{smmnhot1}{rgb}{0,0.5,0}
\definecolor{smmnhot2}{rgb}{1.0,0.5,0.5}
\definecolor{smmnhot3}{rgb}{0.5,0.0,0.0}
\definecolor{ablalight}{rgb}{0.9,0.9,0.0}
\definecolor{ablaimf}{rgb}{0.0,0.25,0.75}
\definecolor{ablamf}{rgb}{1.0,0.0,0.0}
\definecolor{gemini0s}{rgb}{0.9,0.9,0.0}
\definecolor{gemini1sh}{rgb}{0.0,0.0,0.5}
\definecolor{gemini1sl}{rgb}{0.0,0.5,1.0}
\definecolor{gemini2s}{rgb}{0.0,1.0,1.0}

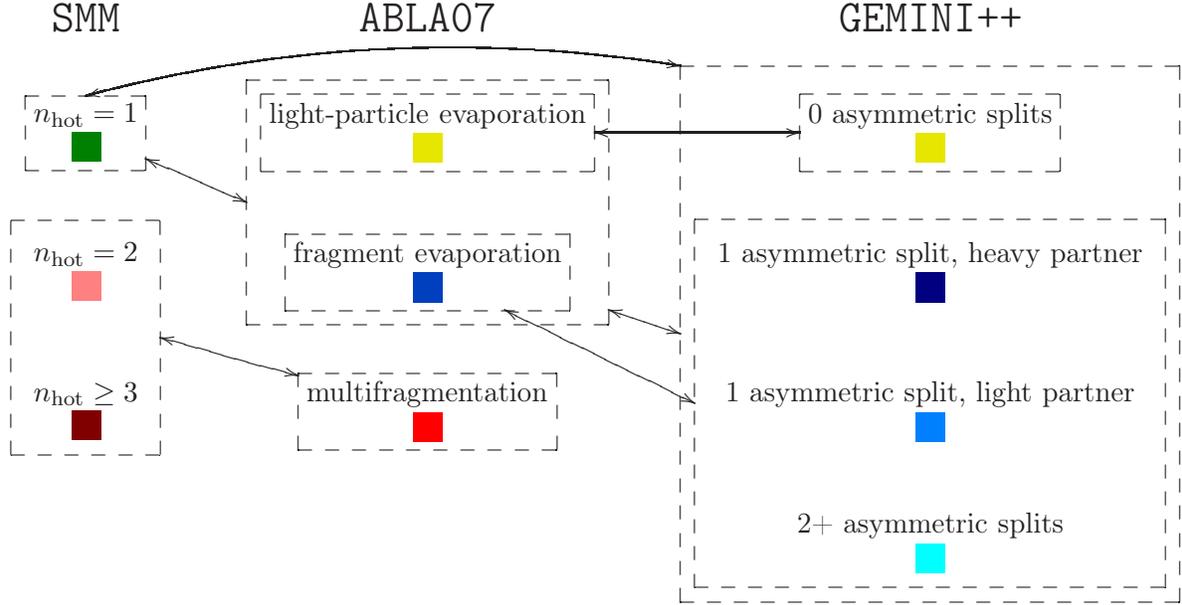
\begin{figure}
  \[
  \xymatrix@C=4em{
    *\txt{\Large\smm} & *\txt{\Large\abla} & *\txt{\Large\geminipp}\\
    *+[F--]\txt{$n_\text{hot}=1$\\\textcolor{smmnhot1}{\rule[-0.5ex]{1em}{1em}}}&
    *+[F--]\txt{light-particle evaporation\\\textcolor{ablalight}{\rule[-0.5ex]{1em}{1em}}}\save[].[d]!C="ablanomf"*+<1em>[F--]\frm{}\restore&
    *+[F--]\txt{0 asymmetric splits\\\textcolor{gemini0s}{\rule[-0.5ex]{1em}{1em}}}\save[].[d].[dd].[ddd]!C="geminiall"*+<2em>[F--]\frm{}\restore\\
    \save[].[d]!C="smmmf"*+<1em>[F--]\frm{}\restore\txt{$n_\text{hot}=2$\\\textcolor{smmnhot2}{\rule[-0.5ex]{1em}{1em}}}&
    *+[F--]\txt{fragment evaporation\\\textcolor{ablaimf}{\rule[-0.5ex]{1em}{1em}}}&
    \save[].[dd]!C="geminiimf"*+<1em>[F--]\frm{}\restore\txt{1 asymmetric split, heavy partner\\\textcolor{gemini1sh}{\rule[-0.5ex]{1em}{1em}}}\\
    *\txt{$n_\text{hot}\geq3$\\\textcolor{smmnhot3}{\rule[-0.5ex]{1em}{1em}}}&
    *+[F--]\txt{multifragmentation\\\textcolor{ablamf}{\rule[-0.5ex]{1em}{1em}}}&
    *\txt{1 asymmetric split, light partner\\\textcolor{gemini1sl}{\rule[-0.5ex]{1em}{1em}}}\\
    &
    &
    *\txt{2+ asymmetric splits\\\textcolor{gemini2s}{\rule[-0.5ex]{1em}{1em}}}
    \ar @{<->} "ablanomf"+L+/l 0.5em/ ; "2,1"
    \ar @{<->} "ablanomf"+DR+/r 0.5em/ ; "geminiall"+L+/l 1em/
    \ar @{<->} "smmmf"+R+/r 0.5em/ ;"4,2"
    \ar @{<->} "geminiimf"+L+/l 0.5em/ ;"3,2"
    \ar @{<->} @/_1.1em/ "geminiall"+UL+/ul 1.414em/ ;"2,1"+U
    \ar @{<->} "2,2" ; "2,3"
  }
  \]
  \caption{Summary of the partitioning of production mechanisms in the different de-excitation models. Equivalent
    mechanisms are connected by arrows and are represented by the same or similar colours in
    Figs.~\ref{fig:p_smm_mechanism}--\ref{fig:p_geminipp_mechanism} and
    \ref{fig:velocity_fe_components_selection}--\ref{fig:velocity_xe_components_selection}. See text for more details
    about the partitioning.}
  \label{fig:mechanisms}
\end{figure}

The analysis of the production mechanism will focus on the coupling of the de-excitation models with \incl, but the
results are qualitatively valid for \isabel, too.

\begin{figure}
  \centering
  \includegraphics[height=0.34\linewidth]{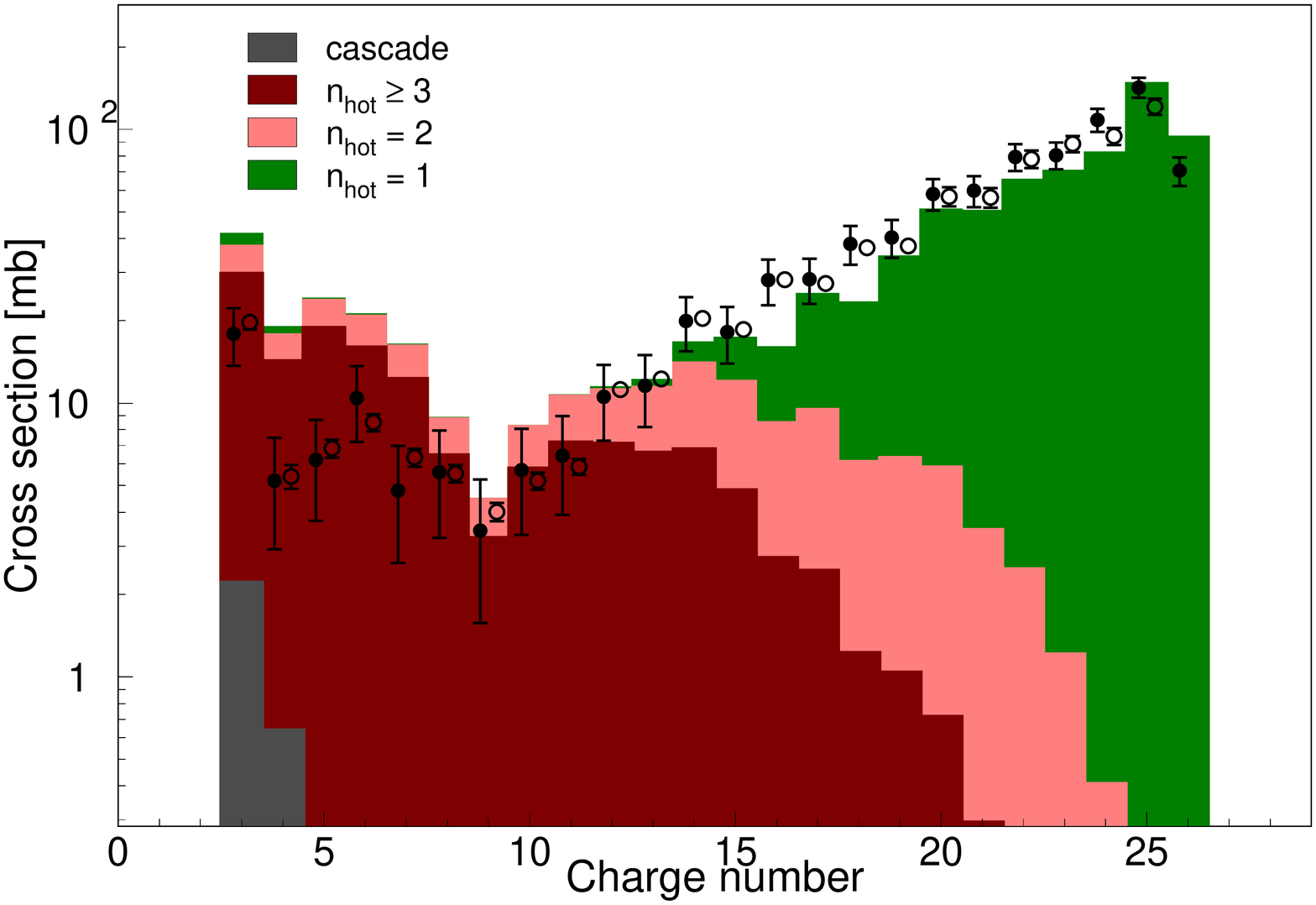}\,\,%
  \includegraphics[height=0.34\linewidth]{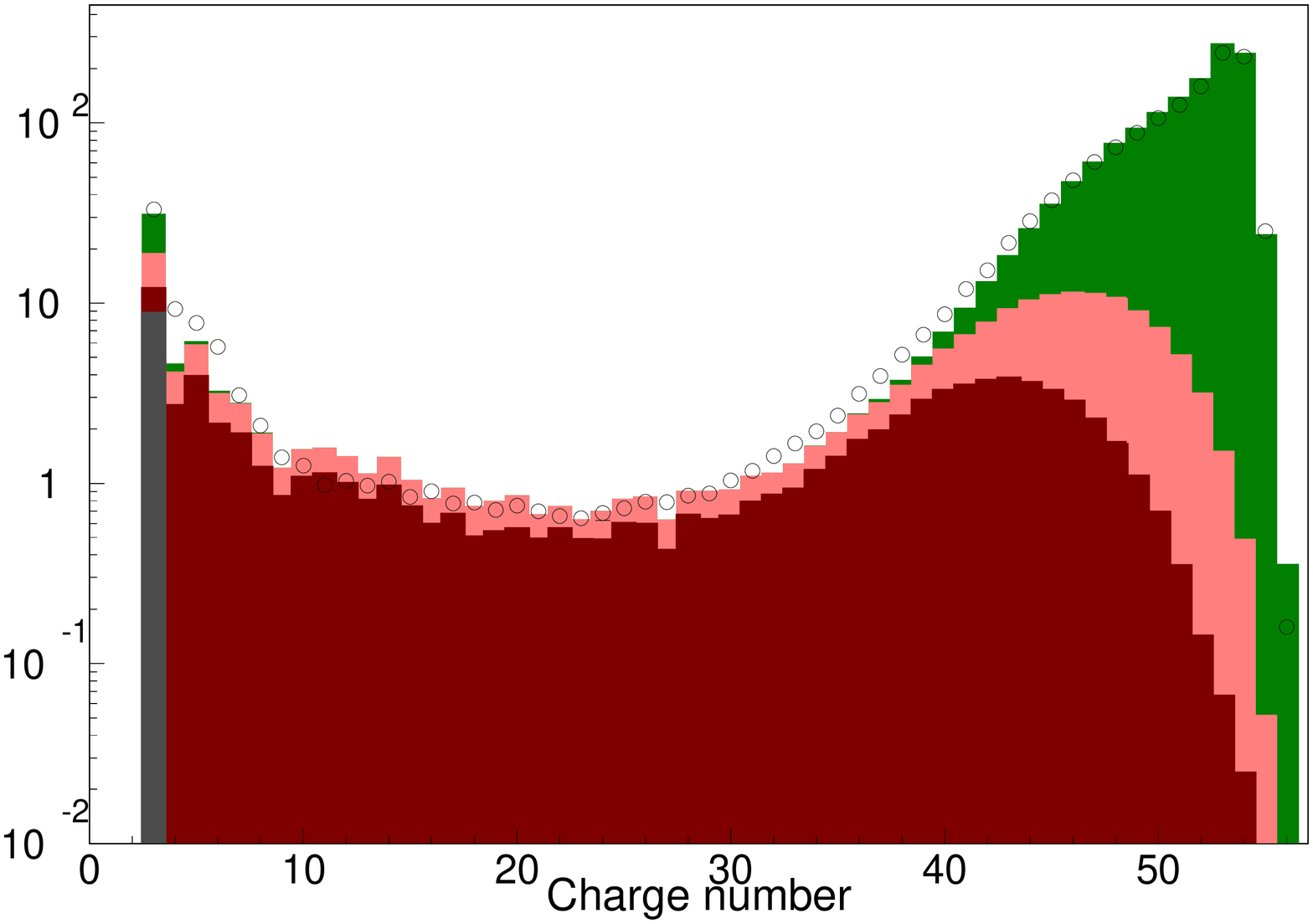}
  \caption{Decomposition of the residue-production cross sections predicted by \incl/\smm\ according to the
    number of hot fragments produced in the multifragmentation stage ($n_\text{hot}$). Left panel: \proton+\iron. Right
    panel: \proton+\xenon.  Experimental data from
    Refs.~\onlinecite{legentil-fe,villagrasa-fe,napolitani-fe,napolitani-xe_xsec}.}
  \label{fig:p_smm_mechanism}
\end{figure}

We start by analysing \smm\ (Fig.~\ref{fig:p_smm_mechanism}). Firstly, we identify the cascade component of the cross
section. The rest of the residue-production cross section is partitioned by labelling each simulated event with the number
$n_\text{hot}$ of hot fragments that emerge from \smm's initial multifragmentation sub-module (and that later de-excite by
sequential evaporation).

Fig.~\ref{fig:p_smm_mechanism} suggests that the plateau cross sections in \proton+\xenon\ and the IMF
cross sections in \proton+\iron\ are almost entirely due to multifragmentation. However, care must be exercised with this
definition of the multifragmentation contribution. Firstly, the onset of
multifragmentation in \smm\ is smooth. Close to the multifragmentation threshold, the most probable break-up configuration
is binary, with one break-up partner much larger than the other. Such processes are similar to (and probably
indistinguishable from) binary decays, and somehow provide a smooth transition to the real multifragmentation
regime. Moreover, \smm\ can produce events where composite fragments are evaporated during the secondary de-excitation of
the hot fragments. Whether such events should be counted as multifragmentation is unclear. In our analysis, these events
are simply classified according to the multiplicity of hot multifragmentation products. Thus, the importance of
multifragmentation for fragment production cannot be easily extracted from the partitioning in
Fig.~\ref{fig:p_smm_mechanism} and would require better event labelling, which is unfortunately unavailable at the
moment. The predicted cross sections for nominal multifragmentation ($n_\text{hot}\geq2$), which can be interpreted as
upper limits for the ``real'' multifragmentation cross section, are 146.0~mb (\iron) and 137.7~mb (\xenon), which
correspond to 18.7\% and 10.0\% of the respective reaction cross sections.

\begin{figure}
  \centering
  \includegraphics[height=0.34\linewidth]{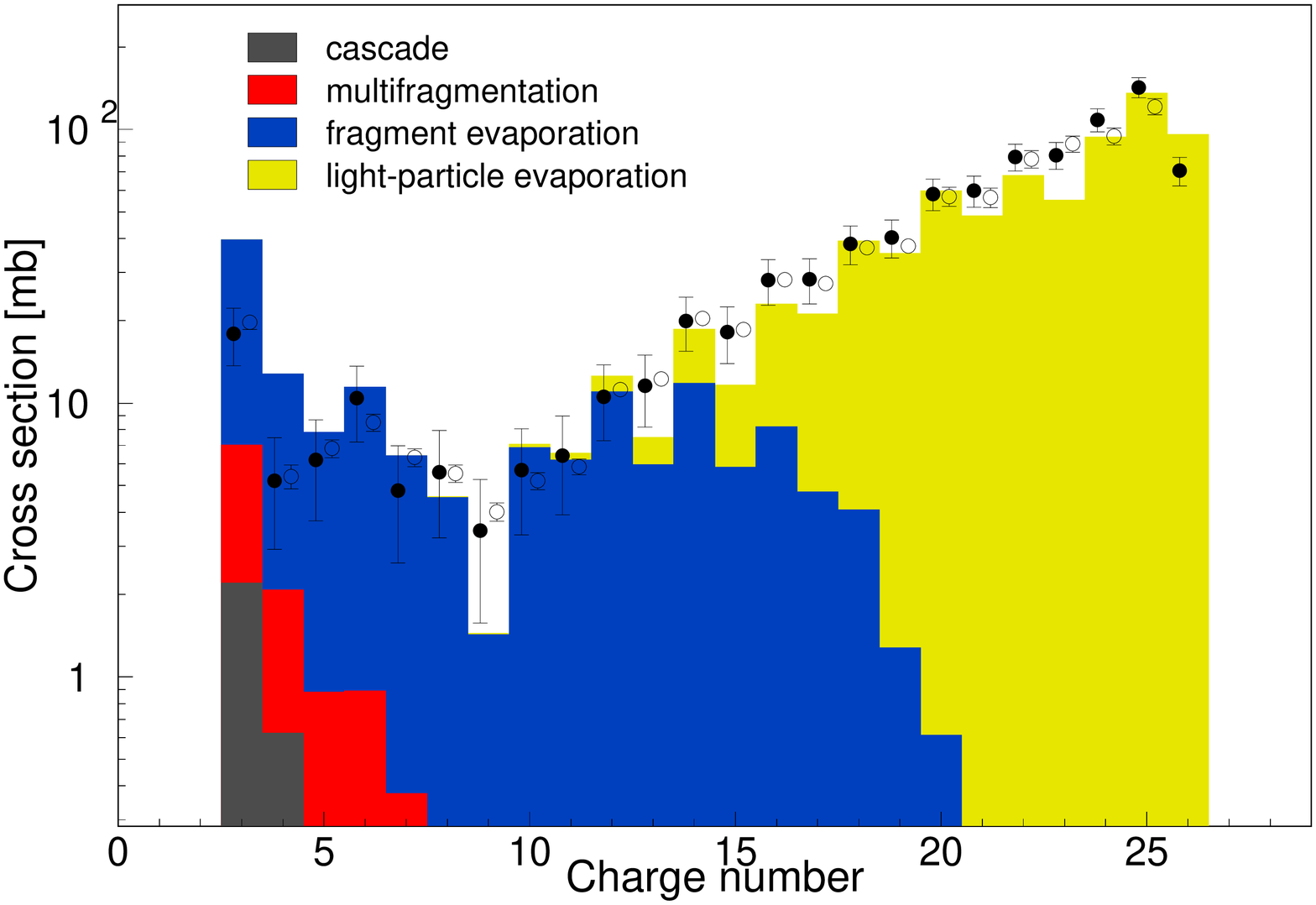}\,\,%
  \includegraphics[height=0.34\linewidth]{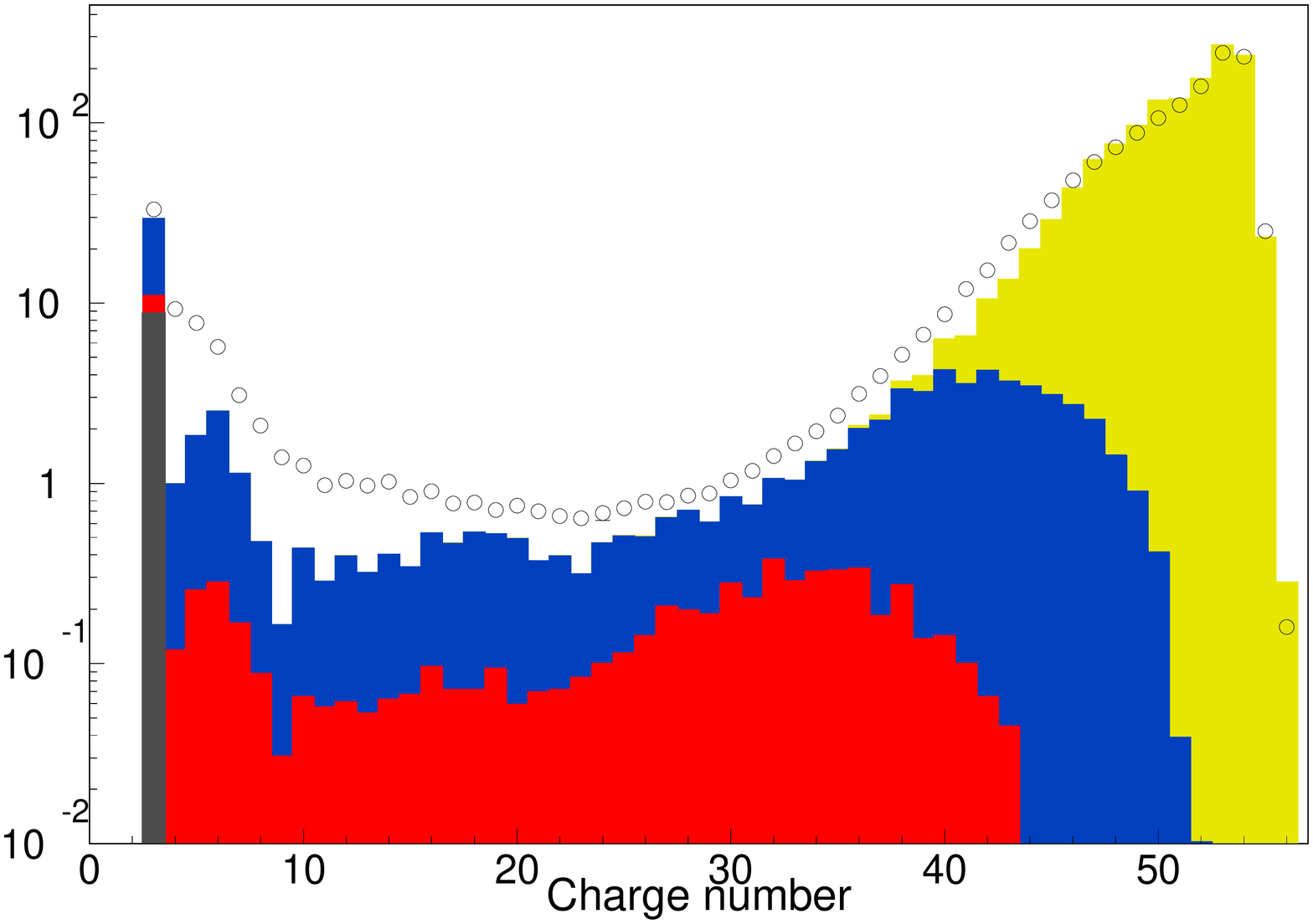}
  \caption{Decomposition of the residue-production cross sections predicted by \incl/\abla\ according to the de-excitation
    mechanism. Left panel: \proton+\iron. Right panel: \proton+\xenon. Experimental data from
    Refs.~\onlinecite{legentil-fe,villagrasa-fe,napolitani-fe,napolitani-xe_xsec}.}
  \label{fig:p_abla07v4_mechanism}
\end{figure}

Figure~\ref{fig:p_abla07v4_mechanism} displays the partitioning of the \incl/\abla\ cross sections. Events that triggered
\abla's multifragmentation module are classified as ``multifragmentation''; as in the case of \smm, this might include
some contamination from secondary fragment evaporation after nominal multifragmentation. Events that did not trigger
multifragmentation are catalogued as ``fragment evaporation'' if one or more fragments were emitted, and as
``light-particle evaporation'' otherwise. Note that events with excitation energies below the particle-emission threshold
are also classified as ``light-particle evaporation''.

One can observe that the cross sections for nominal multifragmentation are much smaller than in the case of \smm: 13.3~mb
(1.7\% of the reaction cross section) for \proton+\iron, and 4.6~mb (0.3\%) for \proton+\xenon. This is
due to \abla's higher multifragmentation threshold. According to Eq.~(\ref{eq:freeze-out}), the freeze-out
temperatures for \iron\ and \xenon\ are 7.97 and 6.36~MeV, respectively, which correspond (assuming Ignatyuk's
level-density parametrisation \cite{Ignatyuk75}) to excitation energies of 6.22 and 3.70~$A$MeV. This should be compared
with the typical \smm\ threshold of 3~$A$MeV. Since the remnant cross section drops fast as the excitation energy
increases, even a moderate difference in the multifragmentation threshold can result in a large cross section difference.

We anyway stress that most of \abla's IMF cross section in \proton+\iron\ does \emph{not} originate from
multifragmentation events. This is at variance with previous claims \cite{legentil-fe}, based on \inclbare\code{4.2}/\gem,
that the \proton+\iron\ residue-production cross sections could not be explained by evaporation.  We stress that, firstly,
\incl/\gem\ provides results similar to \incl/\abla\ on \proton+\iron, and only fails to describe the \proton+\xenon\
data; secondly, \incl/\abla\ is also able to describe the \proton+\xenon\ plateau cross sections mostly thanks to
evaporation alone. Therefore, it is not possible to exclude evaporation solely on the basis of the results of one code
(\gem) for one system (\iron). The details of the evaporation model are obviously important, since \abla\ is able to
provide adequate agreement with all the experimental data considered so far.

\begin{figure}
  \centering
  \includegraphics[height=0.34\linewidth]{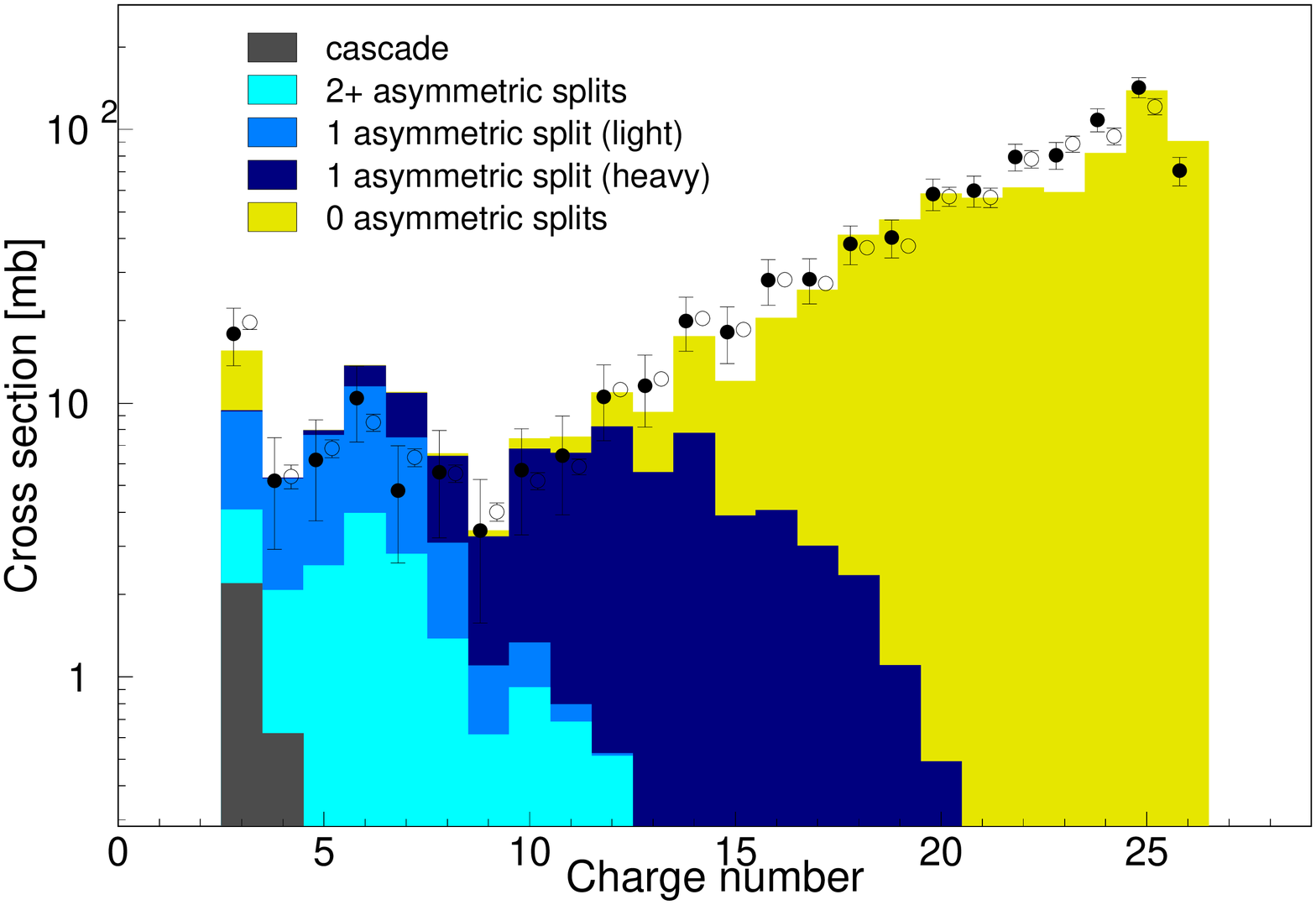}\,\,%
  \includegraphics[height=0.34\linewidth]{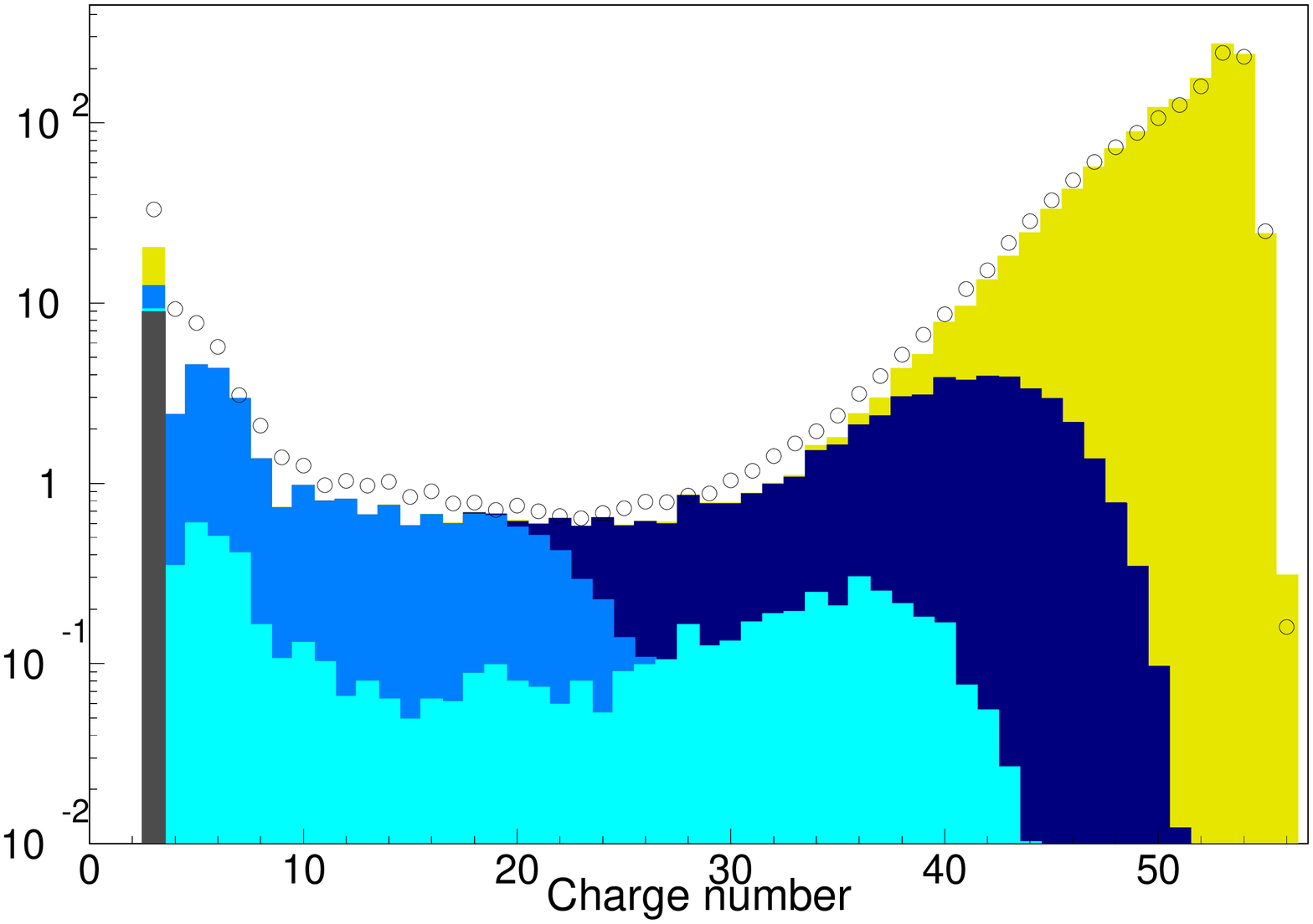}
  \caption{Decomposition of the residue-production cross section predicted by \incl/\geminipp\ according to the number of
    asymmetric splits leading to a given fragment. Left panel: \proton+\iron. Right panel: \proton+\xenon. Experimental
    data from \cite{legentil-fe,villagrasa-fe,napolitani-fe,napolitani-xe_xsec}.}
  \label{fig:p_geminipp_mechanism}
\end{figure}

Finally, the \incl/\geminipp\ partitioning is shown in Fig.~\ref{fig:p_geminipp_mechanism}. De-excitation particles are
classified according to the number of asymmetric splits that led to their production.
For particles following from only one asymmetric split, we distinguish if they originated from the light or the heavy split
partner.
For example, if a remnant splits
into fragments $A$ and $B$, with $A$ larger than $B$, and $A$ subsequently splits into $C$ and $D$, $B$ will be tallied
in the ``1 asymmetric
split (light)'' histogram, while $C$ and $D$ will be counted as ``2+ asymmetric splits''. Light-particle evaporation ($Z\leq3$)
does not influence the count of asymmetric splits.

Note that we assigned the same colour to the \geminipp\ ``0 asymmetric splits'' and the \abla\ ``light-particle
evaporation'' components, suggesting that the two mechanisms are equivalent (see also Fig.~\ref{fig:mechanisms}). However,
emission of Li isotopes in \geminipp\ is described by the Hauser-Feshbach evaporation formalism (Sec.~\ref{sec:geminipp}),
and therefore is not counted as an asymmetric split. In \abla, on the other hand, emission of Li fragments is counted as
``fragment evaporation''. Therefore, the two classes should be considered equivalent for emission of fragments with
$Z\geq4$.

The \incl/\geminipp\ calculations predict a small symmetric-fission component in \proton+\xenon. Although \xenon\ is below
the Businaro-Gallone point, a small number of symmetric-fission event do occur in remnants with high values of spin and of
the $Z^2/A$ ratio. The cross section for such events is only about $5\times10^{-5}$ times the reaction cross section and
it is not visible on the scale of Fig.~\ref{fig:p_geminipp_mechanism}.

\begin{figure}
  \centering
  \includegraphics[height=0.34\linewidth]{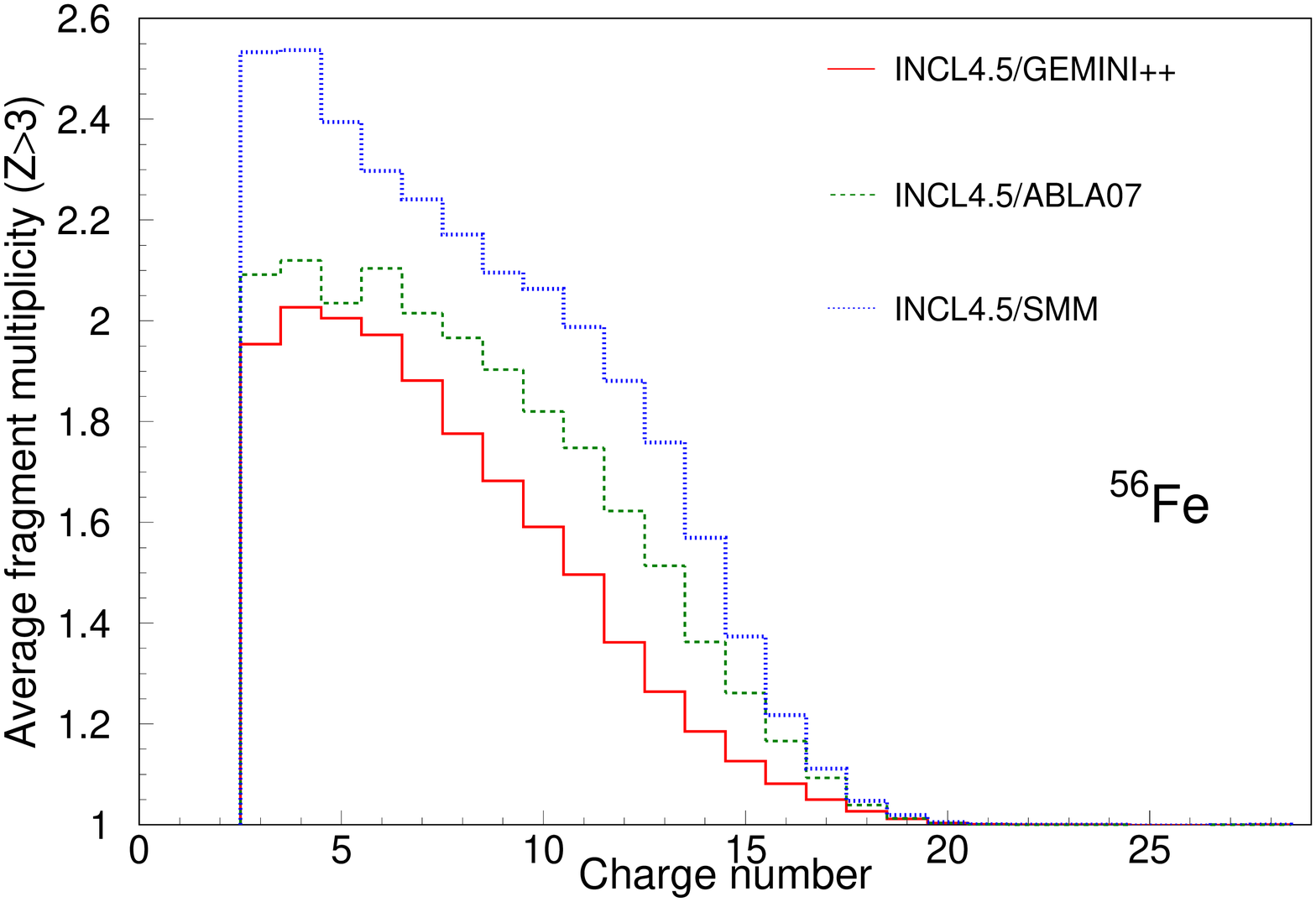}\,\,%
  \includegraphics[height=0.34\linewidth]{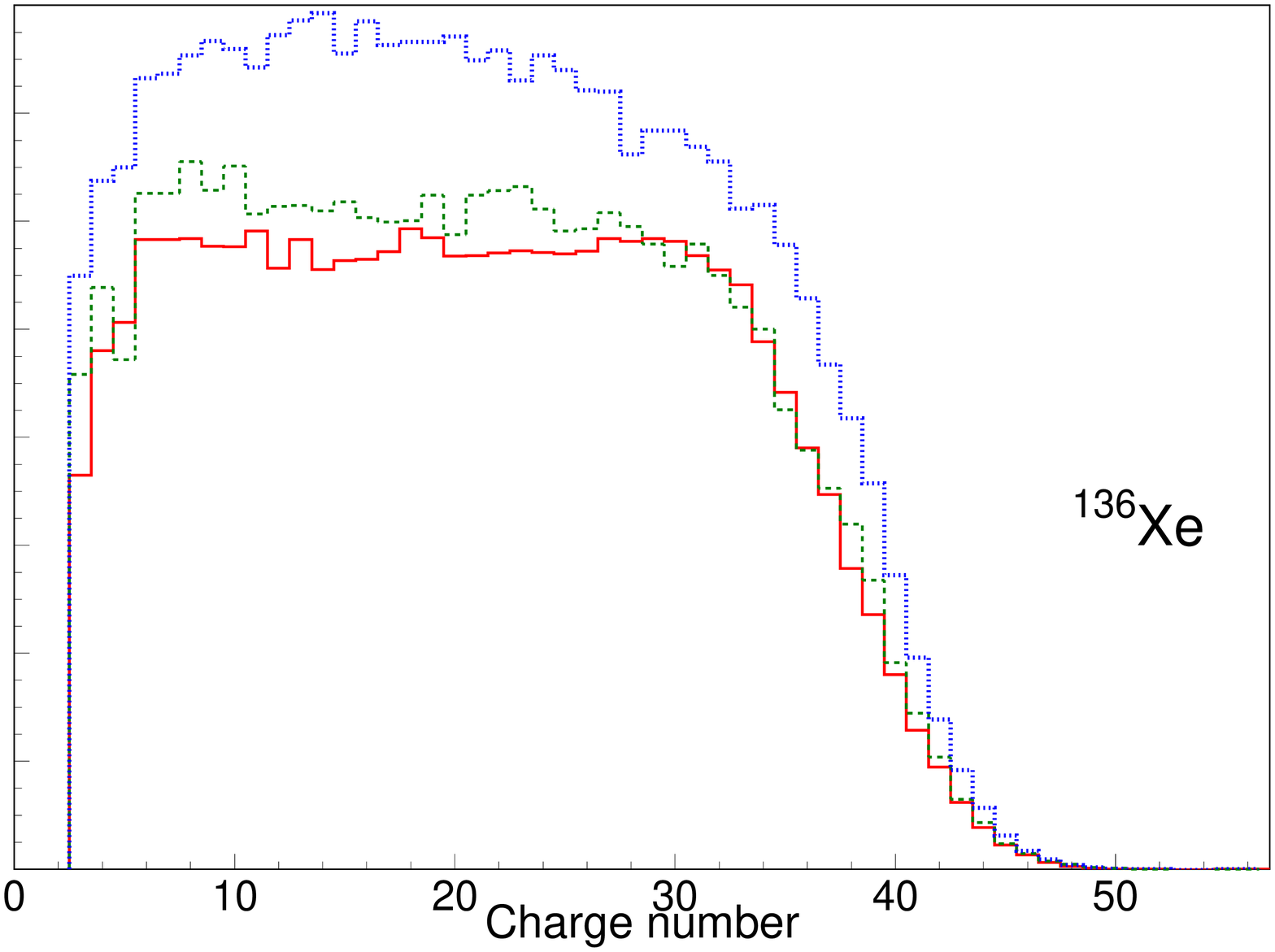}
  \caption{Calculated average multiplicity of particles with $Z\geq3$, as a function of the charge of the fragments
    appearing in the event. See text for more details. Left panel: \proton+\iron. Right panel: \proton+\xenon.}
  \label{fig:average_imf_multiplicity}
\end{figure}

Perhaps unsurprisingly, we find that \geminipp\ generates most of the IMF cross section in \proton+\iron\ and all the
plateau cross section in \proton+\xenon\ through the asymmetric-splitting mechanism. For $Z=3$ the contribution with no
asymmetric split corresponds to evaporated Li nuclei (we remind that emission of nuclei up to $Z=3$ is described by the
evaporation formalism; see Sec.~\ref{sec:geminipp}). There is a striking similarity between the right panels of
Figs.~\ref{fig:p_abla07v4_mechanism} and \ref{fig:p_geminipp_mechanism}. The contributions of nominal multifragmentation
(for \abla) and of pseudo-multifragmentation 2+-split events (for \geminipp) in \proton+\xenon\ are approximately of the
same magnitude and show a similar $Z$ dependence. This suggests that \abla\ and \geminipp\ also predict similar fragment
multiplicities in \proton+\xenon, which is indeed shown to be the case in Fig.~\ref{fig:average_imf_multiplicity}. The
curves represent the average number of particles with $Z\geq3$ as a function of the charge of the fragments that appear in
the event. Thus, for example, \incl/\geminipp\ predicts that neon fragments ($Z=10$) from \proton+\xenon\ appear in events
with on average $\sim2.1$ particles with $Z\geq3$ (including the Ne fragment itself). Interestingly, \incl/\geminipp\ and
\incl/\smm\ predict quite different average fragment multiplicities in the \xenon\ plateau region, despite the
residue-production cross sections being very similar. In general, the average multiplicity does not seem to strictly
correlate with the residue-production cross section. This finding manifestly calls for more exclusive observables, such as
multiplicity distributions and fragment correlations. Such data do exist for \proton+\iron\ (see the following section);
the analysis of a SPALADIN-type \proton+\xenon\ experiment has recently been completed and will soon be published
\cite{gorbinet-thesis}.

As a final remark, we underline that light charged particles emitted during the cascade stage might play some role in the
determination of cross section for the lowest values of $Z$. By default, \incl\ only produces clusters with $A\leq8$,
$Z\leq5$.  The cascade contribution drops off at $Z=4$ because none of the selected $Z=5$ isotopes has $A\leq8$ (see the
beginning of Sec.~\ref{sec:incl-cross-sect} for the discussion on the FRS isotope selection). Almost 50\% of the $Z=3$
cross section in \proton+\xenon\ comes from cascade $^{6,7}$Li.  It is not clear whether heavier clusters might
significantly contribute to the $Z\geq4$ cross sections, which seem to be slightly underestimated by the models.

\section{SPALADIN correlations}\label{sec:spal-corr}

We now turn to the analysis of the model predictions for the SPALADIN \proton+\iron\ data-set \cite{legentil-fe}. Adequate
reproduction of inclusive observables is a prerequisite for the study of semi-exclusive correlations and/or multiplicity
distributions. Therefore, in what follows we will only retain the \incl\ cascade model, which gives residue-production
cross sections that are in better agreement with the experimental data (see Sec.~\ref{sec:incl-cross-sect}).

The goal of the SPALADIN experiment was to measure observables in coincidence for the 1-GeV \proton+\iron\ reaction in
inverse kinematics. These semi-exclusive measurements were obtained at the price of a more sophisticated setup and
experimental analysis than typical inclusive experiments. In Ref.~\onlinecite{legentil-fe} events generated by
cascade/de-excitation models were filtered through a \geant\ transport calculation and analysed like the experimental
data, providing little evidence of convincing multifragmentation signatures.

Since cascade and de-excitation models are in continuous evolution, it is necessary to periodically verify their
predictions on sensitive semi-exclusive observables. The \incl\ and \geminipp\ codes, for example, have largely evolved
since the publication of Ref.~\onlinecite{legentil-fe}. Moreover, the old version of the \code{ABLA} code did not allow
emission of any nucleus heavier than alpha particles, was thus unsuccessful at reproducing inclusive residue-production
cross sections and was therefore excluded from the study of correlations in Ref.~\onlinecite{legentil-fe}. The new \abla\
code, as proven above, can adequately describe the residue-production cross sections; it is therefore interesting to test
its predictions for the SPALADIN correlations. Such considerations provide the motivation for this section.

\begin{figure}
  \centering
  \includegraphics[width=0.6\linewidth]{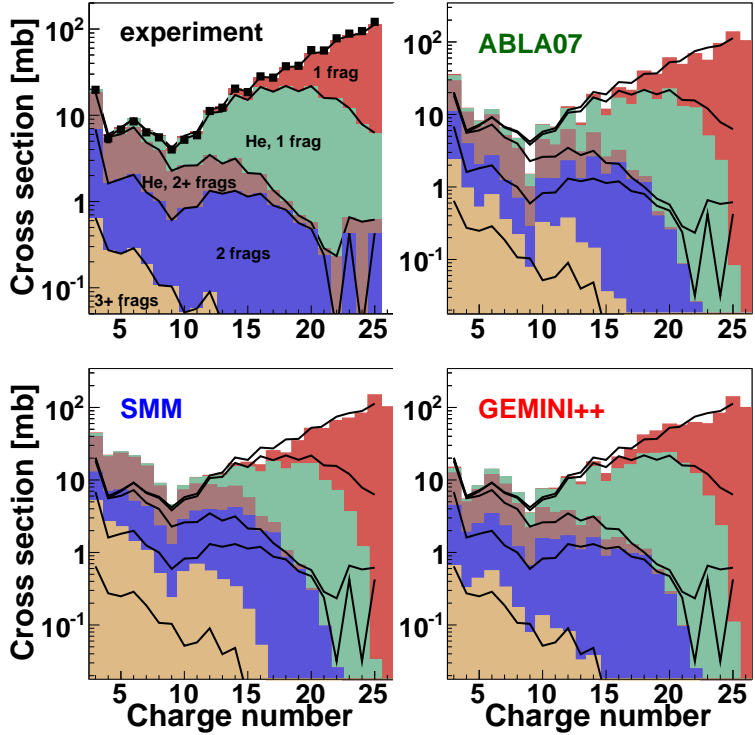}
  \caption{Partitioning of the 1-GeV \proton+\iron\ cross section according to the number of observed helium nuclei and
    IMFs. Top left panel: experimental data. Other panels: solid lines represent the experimental data, histograms
    represent model predictions. The cascade model is \incl. Experimental data from Ref.~\onlinecite{legentil-fe}.}
  \label{fig:spaladin_decomposition}
\end{figure}

We start by analysing the decomposition of the residue-production cross sections. Experimental events were subdivided into
five classes, according to the number of fragments ($Z\geq3$) and helium nuclei that were detected. Modelled events were
run through a \geant\ filter that reproduces the experimental setup and simulates detector efficiency, and subsequently
categorised just like the experimental events. Figure~\ref{fig:spaladin_decomposition} displays the result of this
exercise. Clearly \incl/\geminipp\ and \incl/\abla\ yield the best description of the experimental data. The appraisal of
\smm's results in the IMF region appears to be biased by its slight overestimation of the residue-production cross
sections. Note also that \smm's and \abla's overestimation of 3+-fragment events, which are properly reproduced by
\geminipp, follow the same trend as the average fragment multiplicities depicted in
Fig.~\ref{fig:average_imf_multiplicity}. The \incl/\smm\ and \incl/\geminipp\ results are qualitatively similar to those
presented in Ref.~\onlinecite{legentil-fe}, although both the cascade model and the de-excitation models are sensibly
different.

\begin{table}
  \centering
  \begin{tabular}{c||D{,}{\pm}{3.3}|D{,}{\pm}{3.3}|D{,}{\pm}{3.3}}
    \multirow{2}{*}{model} & \multicolumn{3}{c}{$E^*/A_\text{remnant}$ (MeV)}\\
    & \multicolumn{1}{c|}{$1\leq M_\text{n}+M_\text{He}<3$} & \multicolumn{1}{c|}{$3\leq M_\text{n}+M_\text{He}<5$} &
    \multicolumn{1}{c}{$5\leq M_\text{n}+M_\text{He}<7$}\\
    \hline
    \incl/\smm&4.4,1.3 &4.8,1.4 &5.3,1.4 \\
    \incl/\abla&4.4,1.1 &4.7,1.1&5.2,1.1 \\
    \incl/\geminipp&4.2,1.5&4.9,1.5 &5.6,1.5 \\
    \inclbare\code{4.2}\footnote{Considered independent of the de-excitation model in
      Ref~\onlinecite{legentil-fe}. Root-mean-square values were not provided. See text, Ref.~\onlinecite{legentil-fe} and
      Ref.~\onlinecite{legentil-phd} for more
      details.}&\multicolumn{1}{c|}{3.1}&\multicolumn{1}{c|}{3.8}&\multicolumn{1}{c}{4.5}
  \end{tabular}
  \caption{Remnant excitation energies per nucleon in 1-GeV \proton+\iron\ in events with two detected
    fragments with $Z\geq3$, for three bins in detected neutron-plus-helium multiplicity, as calculated by the three
    de-excitation models used in this work (coupled to \incl) and by the \inclbare\code{4.2} calculations performed by Le
    Gentil \etal\ \cite{legentil-fe}. The values should be interpreted as mean value $\pm$ root mean square.}
  \label{tab:average_estar_mult}
\end{table}

We now address correlations between fragment charges. We restrict our attention to events with at least two detected
fragments ($Z\geq3$) and we define $Z_1$ and $Z_2$ to be the largest and the second-largest observed charges. For these
events, we define three bins in detected neutron-plus-helium multiplicity (1--2, 3--4 and 5--6), which is expected to be
fairly correlated with the excitation energy of the cascade remnant. The correlation between detected neutron-plus-helium
multiplicity and excitation energy was studied with \inclbare\code{4.2} and was shown to be essentially independent of the
de-excitation model \cite{legentil-phd}; however, our results contradict this
conclusion. Table~\ref{tab:average_estar_mult} shows the average excitation energies per nucleon in events with two
$(Z\geq3)$ fragments as a function of the neutron-plus-helium multiplicity. For comparison, we also provide the values
computed by Le Gentil \etal\ with \inclbare\code{4.2} \cite{legentil-fe}, which were claimed to be independent of the
de-excitation model \cite{legentil-phd}. Firstly, we observe that the average excitation energies are \emph{not}
independent of the de-excitation model, at least for the high-multiplicity bin. Note however that the distributions of
excitation energies within each bin are broader than the differences in average excitation energies among neighbouring
bins; thus, the detected neutron-plus-helium multiplicity cannot be interpreted as a \emph{precise} measure of the remnant
excitation energy.  Secondly, the average excitation energies that we find with \incl\ are consistently higher than those
determined by Le Gentil \etal\ using \inclbare\code{4.2}. Thus, the correlation between neutron-plus-helium multiplicity
and excitation energy depends at least on the cascade model; neutron-plus-helium multiplicities cannot represent a
universal, model-independent measure of the remnant excitation energy.

\begin{figure}
  \centering
  \includegraphics[height=0.34\linewidth]{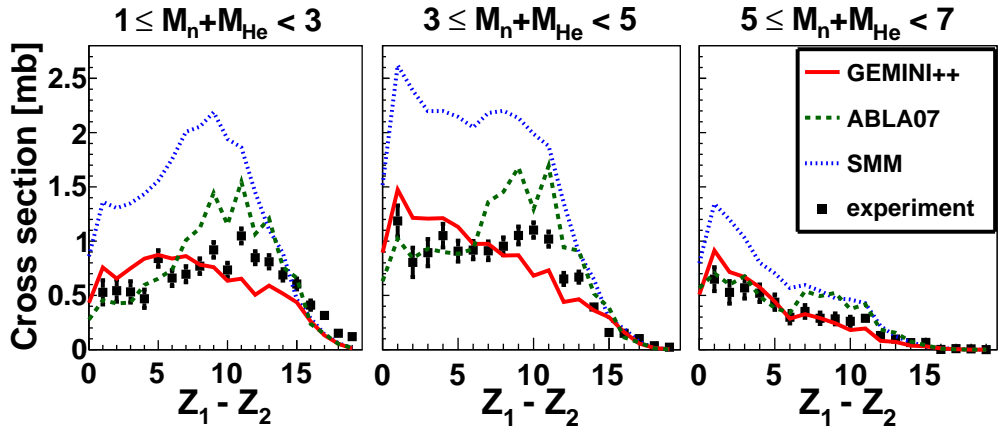}
  \caption{Distributions of $Z_1-Z_2$ (see text) for three different bins in neutron-plus-helium multiplicity, in events
    with at least two fragments with $Z\geq3$. See the text for more details about the excitation energies sampled in
    these bins. The cascade model is \incl. Experimental data from Ref.~\onlinecite{legentil-fe}.}
  \label{fig:spaladin_z1z2}
\end{figure}

Figure~\ref{fig:spaladin_z1z2} shows the distributions of the $Z_1-Z_2$ difference in the three multiplicity bins. As
explained in Ref.~\onlinecite{legentil-fe}, the experimental data indicate that asymmetric ($Z_1\gg Z_2$) and symmetric
($Z_1\simeq Z_2$) charge configurations are favoured at low and high excitation energy, respectively. All the models
reflect this qualitative trend, although only \geminipp\ and \abla\ are able to reproduce the absolute cross sections with
good accuracy. \smm's overestimation is qualitatively consistent with its predictions of IMF yields; in fact, we observe
that the cross sections calculated with a given model, if summed over $Z_1-Z_2$ and over the multiplicity bins, are
quantitatively comparable to the IMF-production cross sections shown in Fig.~\ref{fig:p_fe_inclusive}.

Compared to Le Gentil \etal's results, our work confirms that \incl/\geminipp\ provides the best description of the
$Z_1-Z_2$ distributions. The new \abla\ model, when coupled with \incl, also provides a very good reproduction of the
experimental data. However, one should also remark that the shapes of the \incl/\smm\ distributions are now quite similar
to the experimental data, which was not the case in Ref.~\onlinecite{legentil-fe}, and only the normalisation seems to be
consistently off by a factor of about two. This is quantitatively consistent with \incl/\smm's overestimation of the
residue-production cross sections (Fig.~\ref{fig:p_fe_inclusive}) and raises an interesting question, i.e.\ whether it
might be possible to adjust \incl/\smm\ to better reproduce residue-production cross sections and the SPALADIN observables
at the same time.

\begin{figure}
  \centering
  \includegraphics[width=0.7\linewidth]{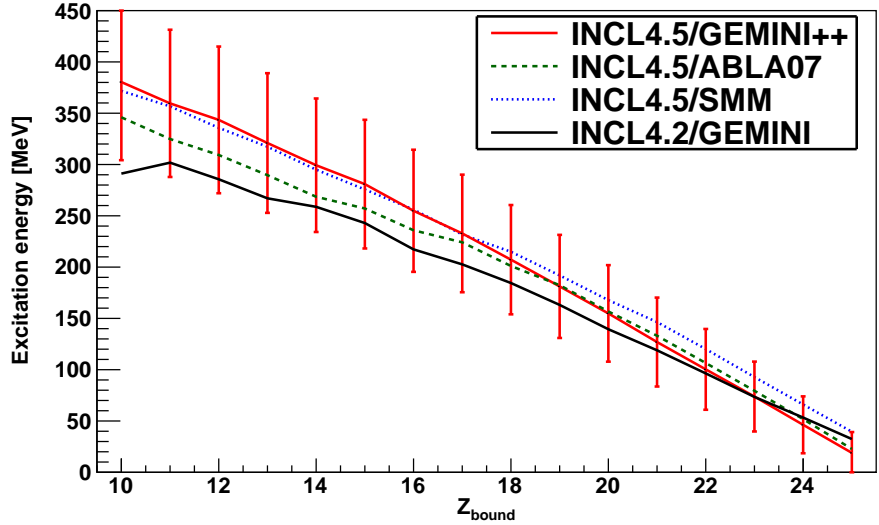}
  \caption{Average remnant excitation energy as a function of the value of $Z_\text{bound}$, which is defined as the sum
    of all the detected charges greater than one. The error bars represent the width of the excitation-energy
    distributions for \incl/\geminipp. The black line was extracted from Ref.~\onlinecite{legentil-fe} and was computed
    with \inclbare\code{4.2}/\code{GEMINI}.}
  \label{fig:spaladin_zbound-exini}
\end{figure}

We finally turn to fragment multiplicities. We first need to define the $Z_\text{bound}$ variable as the sum of all the
detected charges with $Z\geq2$. This quantity was previously found to be negatively correlated with the excitation energy
of the cascade remnant, and the correlation was found to be independent of the de-excitation model
\cite{legentil-fe}. Fig.~\ref{fig:spaladin_zbound-exini} demonstrates that both these properties stay true when the
calculations are performed with \incl. The $E^*$-$Z_\text{bound}$ correlation, however, is found to be slightly different
for \inclbare\code{4.2} and \incl. This indicates that $Z_\text{bound}$ should not be considered as a universal,
cascade-model-independent measure of the excitation energy, although it seems relatively robust with respect to the choice
of the de-excitation model.

\begin{figure}
  \centering
  \includegraphics[width=0.9\linewidth]{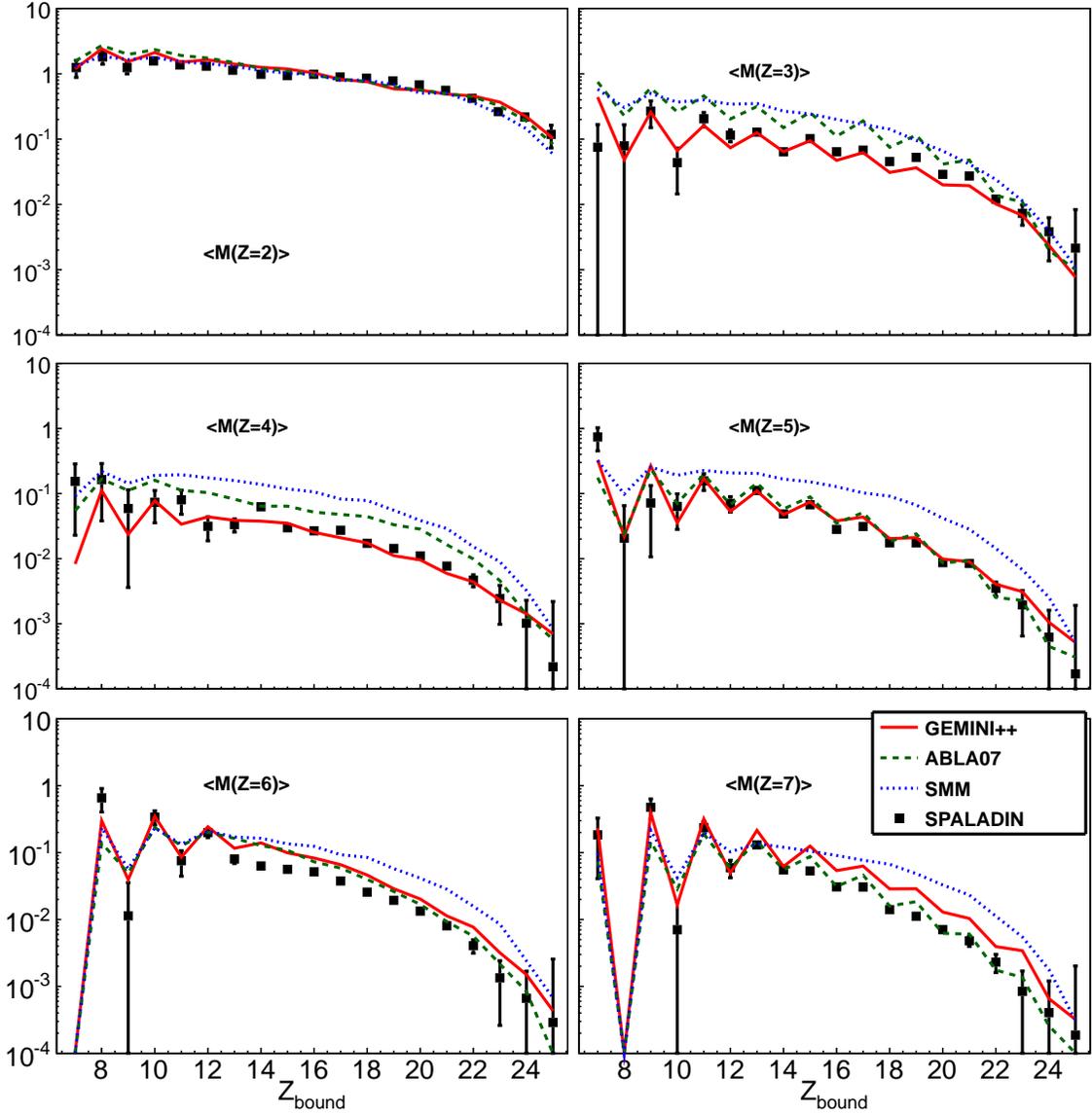}
  \caption{Average fragment multiplicities as a function of the $Z_\text{bound}$ variable (see text). The cascade model is
    \incl. Experimental data from Ref.~\onlinecite{legentil-fe}.}
  \label{fig:spaladin_zbound}
\end{figure}

The average fragment multiplicities (i.e.\ the average number of fragments with a given charge produced in a reaction) are
plotted in Fig.~\ref{fig:spaladin_zbound}, for fragment charges between 2 and 7, as functions of $Z_\text{bound}$. All
multiplicities rise as $Z_\text{bound}$ decreases, reflecting the positive correlation between fragment multiplicity and
excitation energy. Note that production cross sections for $Z=2$ were measured by the SPALADIN collaboration, but they
were not plotted on Fig.~\ref{fig:p_fe_inclusive} because we summed the calculated isotopic cross sections over the
isotopes measured in the FRS experiments (see Sec.~\ref{sec:incl-cross-sect}).

The first striking result is that, for a fixed fragment, all the curves have a very similar shape. If $Z_\text{bound}$ is
interpreted as the excitation energy of the cascade remnant, this indicates that all the de-excitation models predict a
similar dependence of the fragment-emission probability on the excitation energy. The overall level of the curve is
charge-by-charge correlated with the IMF yields in Fig.~\ref{fig:p_fe_inclusive}. Thus, for example, \abla\ and \smm\
predict too large multiplicities and too large production cross sections for $Z=3$. Besides this correlation, it is not
clear whether there is a lesson to be learnt from these observables. The strong even-odd staggering at low
$Z_\text{bound}$ is reproduced by all models.

Summarising, the quest for model-independent measures of the remnant excitation energy is still open. We checked that
neutron-plus-helium multiplicities and the $Z_\text{bound}$ variable are at least sensitive to the cascade model. The new
\geminipp\ and \smm\ versions yield predictions similar to those reported by Ref.~\onlinecite{legentil-fe}, although the
cascade model used in the present work yields rather different excitation-energy and remnant-mass distributions. The
\abla\ and \geminipp\ models can reproduce most of the considered observables. Given the very small fraction of nominal
multifragmentation events predicted by \incl/\abla\ (Sec.~\ref{sec:production-mechanism}), we confirm that explanation of
the SPALADIN data does not require any strong multifragmentation component.

To our knowledge, no published correlation data exist for the \proton+\xenon\ reaction around 1 GeV. However, a
SPALADIN-type experiment was performed in April 2009 and the results of the analysis are due to be published soon
\cite{gorbinet-thesis}.

\section{Longitudinal-velocity distributions}\label{sec:long-veloc-distr}

Insight about the de-excitation mechanism can also be gained by examining the kinematics of the decay products. Sequential
binary splits should produce kinematical patterns reminiscent of the decay barriers; multifragmentation, on the other
hand, is expected to produce fragments with broad, structureless velocity distributions. We discuss here the velocity
distributions measured in the context of the FRS experiments considered in Sec.~\ref{sec:incl-cross-sect}
\cite{napolitani-fe,napolitani-xe_velocity}. The emission velocities of fragments from the 1-GeV \proton+\iron\ and
\proton+\xenon\ reactions are measured using forward spectrometry techniques. Reactions are studied in inverse kinematics,
i.e.\ as a 1-$A$GeV \iron\ or \xenon\ beam impinging on a $^\text{1}$H target. Most of the de-excitation products are
focalised in a cone around the beam axis. The experimental angular acceptance somewhat depends on the trajectory azimuth,
but it is on average equal to 15~mrad. Particles that satisfy the acceptance cut are identified by mass and charge and
their longitudinal velocity (the component of the velocity along the beam axis) is
measured. Refs.~\onlinecite{napolitani-fe} and \onlinecite{napolitani-xe_velocity} report measured longitudinal-velocity
distributions (LVDs) for several nuclides with $A\geq6$.

We used our cascade/de-excitation tools to calculate the longitudinal-velocity distributions for the same nuclides. A
detailed comparison with the experimental LVDs requires knowledge of the azimuthal dependence of the FRS angular
acceptance and a three-dimensional macroscopic transport calculation of the reaction products in the spectrometer. We
limited ourselves to a simpler approach: we assumed the angular acceptance to be independent of the trajectory azimuth and
equal to the experimental average value of 15~mrad. In other words, the acceptance of our simulation is a circular cone in
velocity space, centred on the beam axis, with vertex in the origin and aperture of 15~mrad. This prevents a refined
quantitative comparison of our results with the experimental data, but the emerging trend is nonetheless clear, as we will
show in the following.


\begin{figure}
  \centering
  \includegraphics[width=0.7\linewidth]{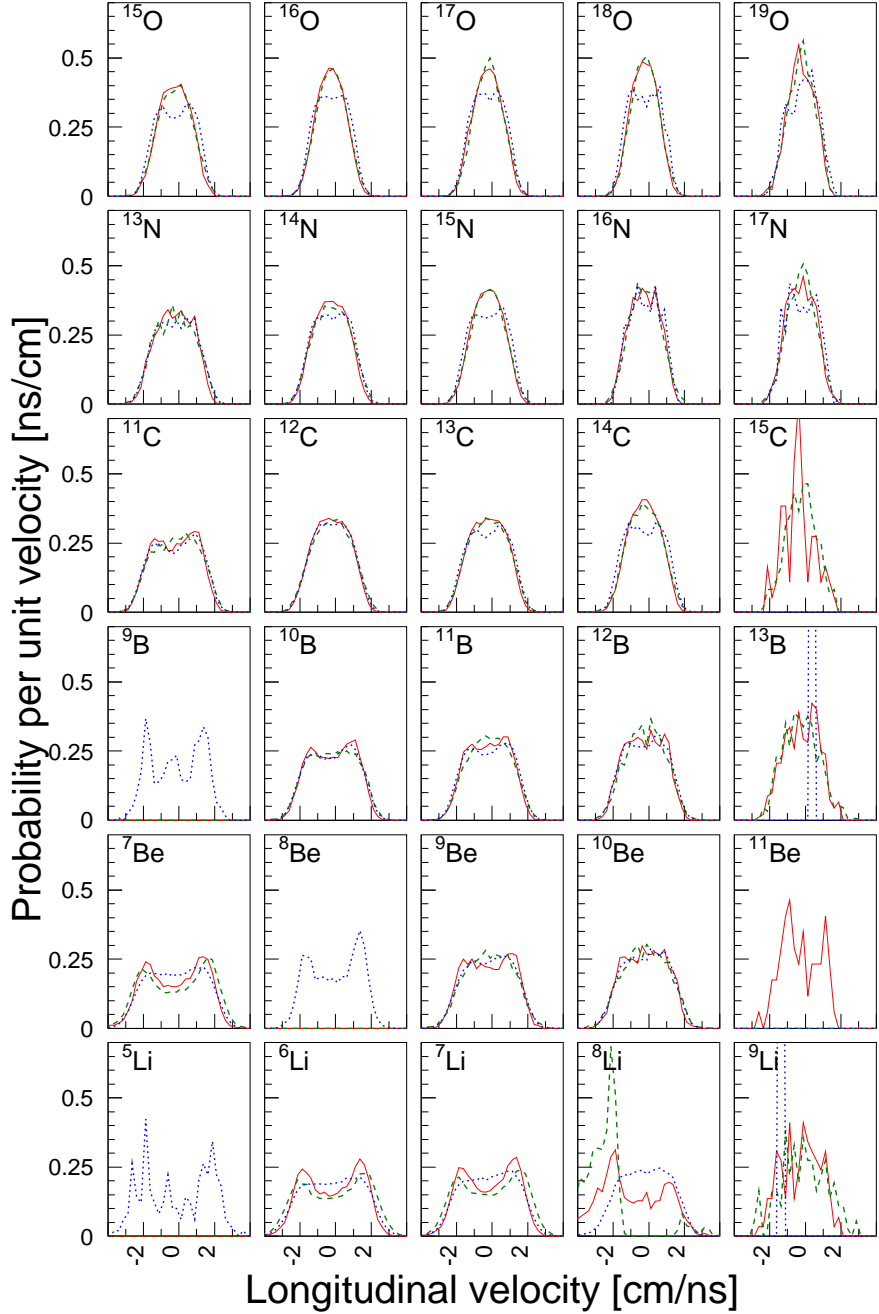}
  \caption{Calculated longitudinal-velocity distributions for 1-$A$GeV \iron+$^\text{1}$H in the beam rest frame. Red
    solid line: \incl/\geminipp; green dashed line: \incl/\abla; blue dotted line: \incl/\smm.}
  \label{fig:velocity_fe}
\end{figure}

\begin{figure}
  \centering
  \includegraphics[width=0.7\linewidth]{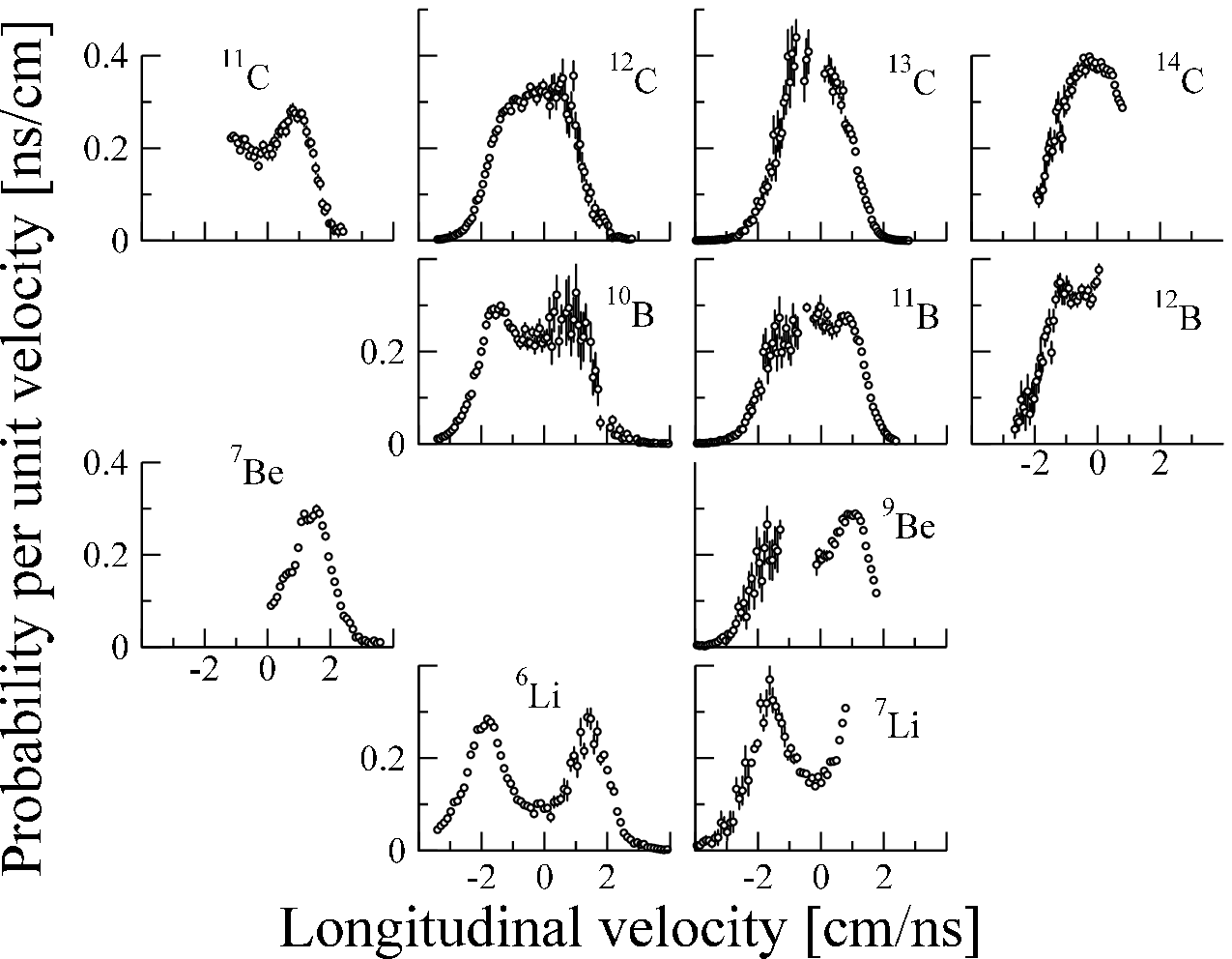}
  \caption{Experimental longitudinal-velocity distributions for 1-$A$GeV \iron+$^\text{1}$H in the beam rest
    frame. Adapted from Ref.~\onlinecite{napolitani-fe}.}
  \label{fig:velocity_fe_exp}
\end{figure}

Figure~\ref{fig:velocity_fe} shows the calculated LVDs for \iron+$^\text{1}$H. Each distribution is separately normalised
to one. Note that, in this and all the following figures, the longitudinal velocities refer to the rest frame of the
\iron\ projectile, with the proton impinging with negative velocity. This choice was made for consistency with the
experimental data plotted in Fig.~\ref{fig:velocity_fe_exp}.

We note that a few short-lived nuclides are present in the \smm\ results ($^\text{5}$Li, $^\text{8}$Be and
$^\text{9}$B). These nuclides would typically decay before being detected by the experimental apparatus. The decays could
in principle populate other IMF species and modify their LVDs, but the nuclides above 
entirely decay in nucleons and alpha particles
. Therefore, we can neglect them in the following discussion.

One observes that all models produce similar, single-peaked distributions for the heaviest IMFs (say for $A\geq9$). Only
for the lightest IMFs can we observe differences among the model predictions, with \geminipp\ and \abla\ often producing
double-peaked distributions, whereas \smm\ typically yields flat distributions.  These predictions should be compared with
the measured distributions \cite[Fig.~10]{napolitani-fe}, which are reported in Fig.~\ref{fig:velocity_fe_exp} for
convenience of the reader. In stark contrast with Napolitani \etal's claims, we find that binary decay does \emph{not}
imply sharp Coulomb holes in the velocity distributions. Indeed, the shapes of the measurements distributions seem to be
best described by \geminipp; compare e.g.\ the double-peaked structure of the $^\text{6,7}$Li distributions, where the
Coulomb peaks predicted by \geminipp\ are possibly even too weak to account for the measured shape.


\begin{figure}
  \centering
  \includegraphics[width=0.9\linewidth]{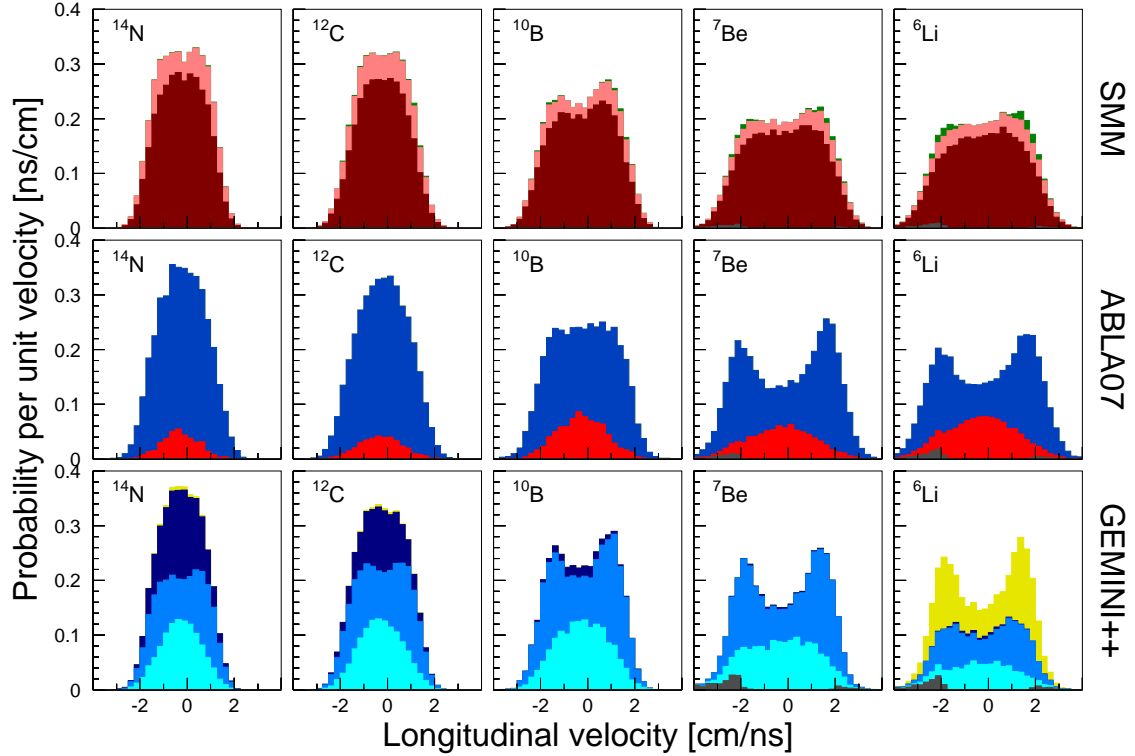}
  \caption{Decomposition of simulated longitudinal-velocity distributions in \iron+$^\text{1}$H for some selected
    nuclides, according to the partitions summarised in Fig.~\ref{fig:mechanisms}.}
  \label{fig:velocity_fe_components_selection}
\end{figure}

How can \geminipp\ produce single-peaked LVDs by relying on its binary-decay mechanism? This question can be answered by
partitioning LVDs according to the production mechanism, as done in Fig.~\ref{fig:velocity_fe_components_selection}.
Firstly, we
observe that nuclei that follow from two or more asymmetric splits (cyan component) expectedly produce single-peaked
distributions.
The ``0 asymmetric splits'' component, which is only present in Li isotopes, corresponds to direct evaporation and shows a
sharp Coulomb hole.  However, contrary to what one would naively expect, nuclei following from one asymmetric split
produce only mildly structured LVDs. Distributions associated with light split partners do show Coulomb holes, although
secondary de-excitation (evaporation of light particles) after the split somewhat blurs the peaks. However, and more
importantly, the heavy partner of asymmetric splits typically picks up very little recoil, but can retain enough
excitation energy to lose much of its mass and be eventually detected as an IMF with small longitudinal velocity. Thus,
a sizable fraction of the single-peaked contribution to the \geminipp\ LVDs in Fig.~\ref{fig:velocity_fe_components_selection} comes
from the ``1 asymmetric split (heavy)'' (dark-blue) mechanisms, which we may term the the \emph{de-excitation-residue
  component}.
This
result somewhat contrasts with the interpretation suggested by \smm, which entirely attributes the single-peaked component
to nominal
multifragmentation events. Finally, observe that the \abla\ ``fragment evaporation'' component (blue), which contains both
evaporated fragments and fragment-evaporation residues, shows no stark Coulomb structure for $A\geq10$, consistently with
the conclusions suggested by \geminipp.

We stress that the participation of de-excitation residues to IMF distributions is possible only because cascade remnants in
\proton+\iron\ are relatively close in $A$ and $Z$ to the IMF mass region (Fig.~\ref{fig:p_remnant}). The residue
component somewhat provides a background noise over which true multifragmentation signatures are superimposed. One of the
motivations for studying the \proton+\xenon\ reaction is exactly that the residue component is expected to be completely
negligible in the IMF region. It is therefore of great interest to consider how well \geminipp\ performs at reproducing
the measured LVDs for this reaction.

\begin{figure}
  \centering
  \includegraphics[width=0.7\linewidth]{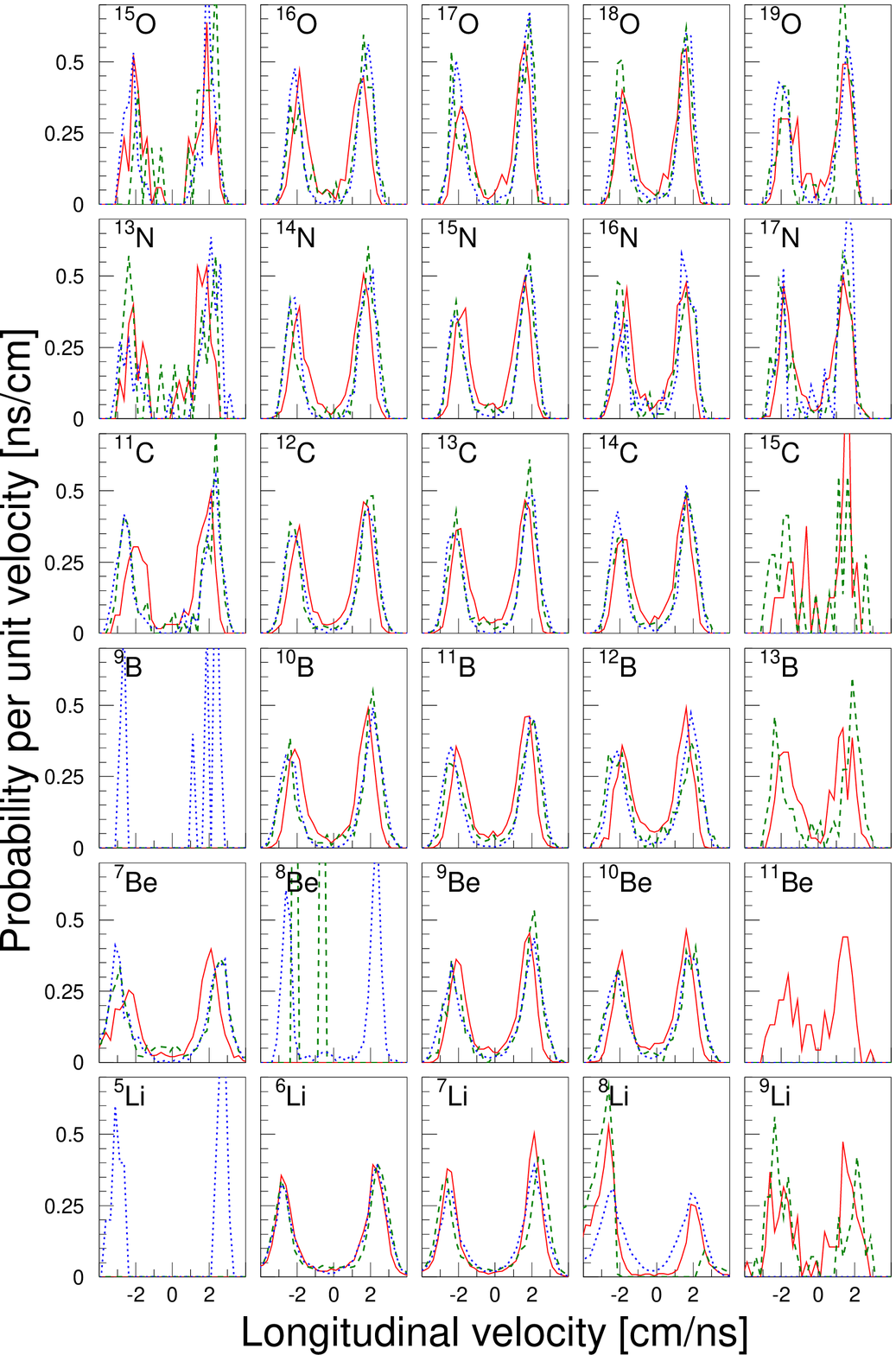}
  \caption{Same as Fig.~\ref{fig:velocity_fe}, for 1-$A$GeV \xenon+$^\text{1}$H.}
  \label{fig:velocity_xe_1}
\end{figure}

\begin{figure}
  \centering
  \includegraphics[width=0.7\linewidth]{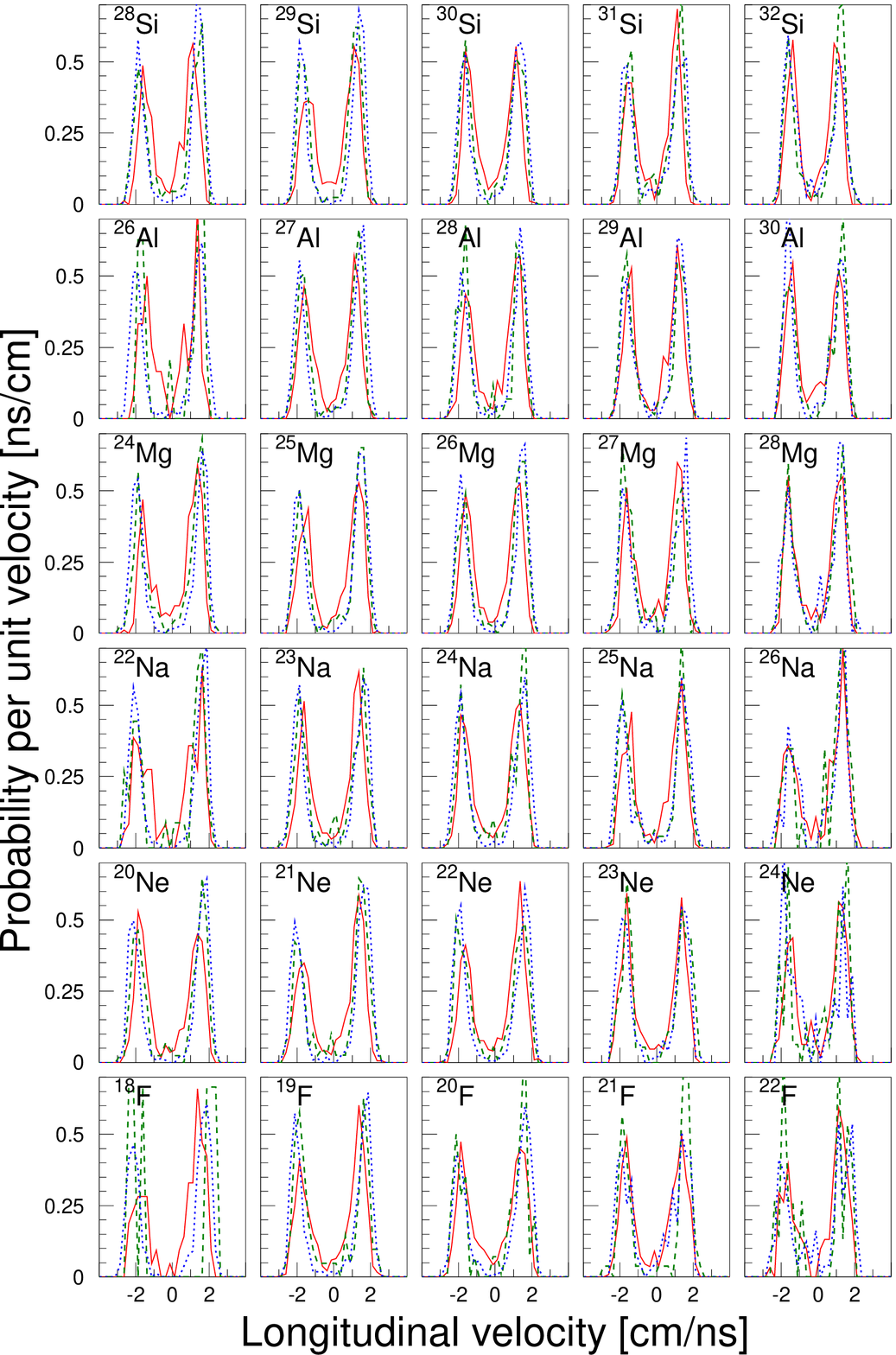}
  \caption{Same as Fig.~\ref{fig:velocity_xe_1}, for $9\leq Z\leq14$.}
  \label{fig:velocity_xe_2}
\end{figure}

As displayed in Figs.~\ref{fig:velocity_xe_1} and \ref{fig:velocity_xe_2}, the LVDs predicted by \geminipp\ show clear
Coulomb holes over all the considered mass and charge range. Indeed, the residue component is negligible and \geminipp\ is
not able to reproduce the experimental distributions \cite[Fig.~2]{napolitani-xe_velocity}, which become rather flat from
$Z\simeq6$, as in the case of \iron. Surprisingly, however, even \abla\ and \smm\ predict double-peaked LVDs that are very
similar to \geminipp's, even for relatively heavy nuclides such as the Si isotopes (Fig.~\ref{fig:velocity_xe_2}). No
model seems to be able to account for the single-peaked component that clearly dominates the experimental LVDs for
$Z\gtrsim6$.

\begin{figure}
  \centering
  \includegraphics[width=0.9\linewidth]{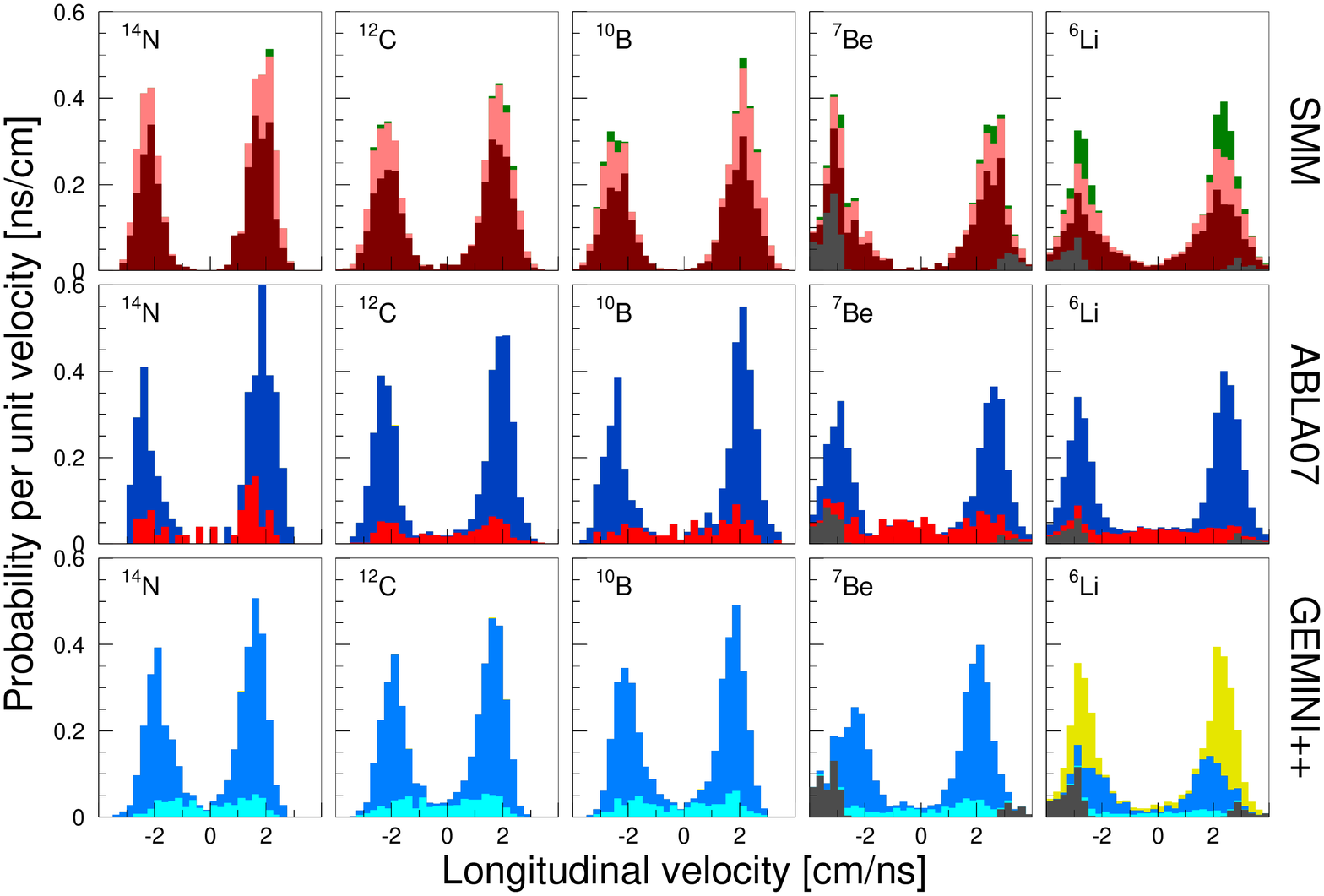}
  \caption{Same as Fig.~\ref{fig:velocity_fe_components_selection}, for \xenon+$^\text{1}$H.}
  \label{fig:velocity_xe_components_selection}
\end{figure}

This unanticipated results clashes with the widespread belief that multifragmentation should yield single-peaked velocity
distributions, especially for high fragment multiplicities. \smm\ accounts for the exact propagation of the hot
multifragmentation products in their mutual Coulomb field, and is therefore supposed to yield reasonable predictions of
the asymptotic velocities, if the initial conditions are realistic. Fig.~\ref{fig:velocity_xe_components_selection}
demonstrates that this mechanism does
\emph{not} yield single-peaked LVDs. Even many-body ($\geq3$) break-up configurations (dark-red component) bear clear signs of
Coulomb repulsion. This suggests that most of the break-up configurations must be quasi-binary, with one or two large
fragments completely dominating the Coulomb dynamics, possibly accompanied by nucleons and very light charged
particles. Note that \abla's multifragmentation mechanism seems to produce flatter distributions, although its cross
section is largely insufficient to explain the experimental shapes.

Finally, we comment briefly about the contribution of dynamical emission of cascade clusters to the LVDs. We remind the
reader that \incl\ by default produces clusters through a coalescence mechanism up to $A=8$ included
\cite{cugnon-incl45_nd2010}. Indeed, Figs.~\ref{fig:velocity_fe_components_selection} and
\ref{fig:velocity_xe_components_selection} show a cascade contribution for the lightest IMFs. Cascade clusters, which are
high-energy particles in direct kinematics, constitute an asymmetric tail that extends in the backward direction in
inverse kinematics. The cascade-cluster tail is sometimes responsible for a large forward-backward asymmetry of the LVD. A
similar signature was observed in the \xenon+$^\text{1}$H data-set \cite{napolitani-xe_velocity}, but was attributed to
fluctuations in the recoil momentum of the cascade remnant. The \incl\ model suggests a different interpretation.

\section{Time interval between fragment emissions}\label{sec:time-interv-betw}

We have so far considered residue-production cross sections, correlations among de-excitation products and distributions
of longitudinal velocity of the emitted fragments. For most of these endpoints, the \incl/\geminipp\ model provided the
most accurate description of the experimental data (with the exception of the \proton+\xenon\ LVDs, which no model seems
to be able to reproduce). This result can be taken as evidence that a multifragmentation model is unnecessary to describe
the reactions studied in the present paper. The solidity of this argument, however, relies on the internal consistency of
the application of the \incl/\geminipp\ model to the systems considered. We will now proceed to show that the time
interval between fragment emissions for highly excited cascade remnants becomes comparable to the typical
multifragmentation timescale.

The \geminipp\ model keeps an internal clock of the decay process, which is readily available to the user. For a compound
nucleus with decay width $\Gamma$, the decay time is sampled from an exponential distribution with time constant
$\hbar/\Gamma$. Thus, we select events with two or more asymmetric splits (\emph{two-split events}) and compute the
interval length $\Delta t_\text{split}$ between the earliest and the second-earliest emission. Note that this definition
is the event-based equivalent of the ``2+ asymmetric splits'' particle classification above
(Sec.~\ref{sec:production-mechanism}). If more than two splits occur during an event, we only consider the two earliest;
the idea is that we are interested in short intervals and emissions become more separated in time as the excitation energy
is evacuated. Note that not every asymmetric split leads to observed fragments; if the excitation energy of the emitted
fragment is sufficient, it can completely disassemble in light charged particles and remain unobserved. Thus, two
asymmetric splits do not necessarily correspond to three observed fragments.

We then define twelve bins in excitation energy per nucleon by requiring that two-split events be uniformly partitioned
over the
bins. Since there are only few two-split events at low excitation energy, the first bin is very broad (from zero to
4.2~$A$MeV for \iron\ and 2.8~$A$MeV for \xenon). Finally, for each bin we construct a distribution of interval
lengths. This procedure permits studying how time intervals evolve as a function of the excitation energy of the cascade
remnant.

\begin{figure}
  \centering
  \includegraphics[height=0.34\linewidth]{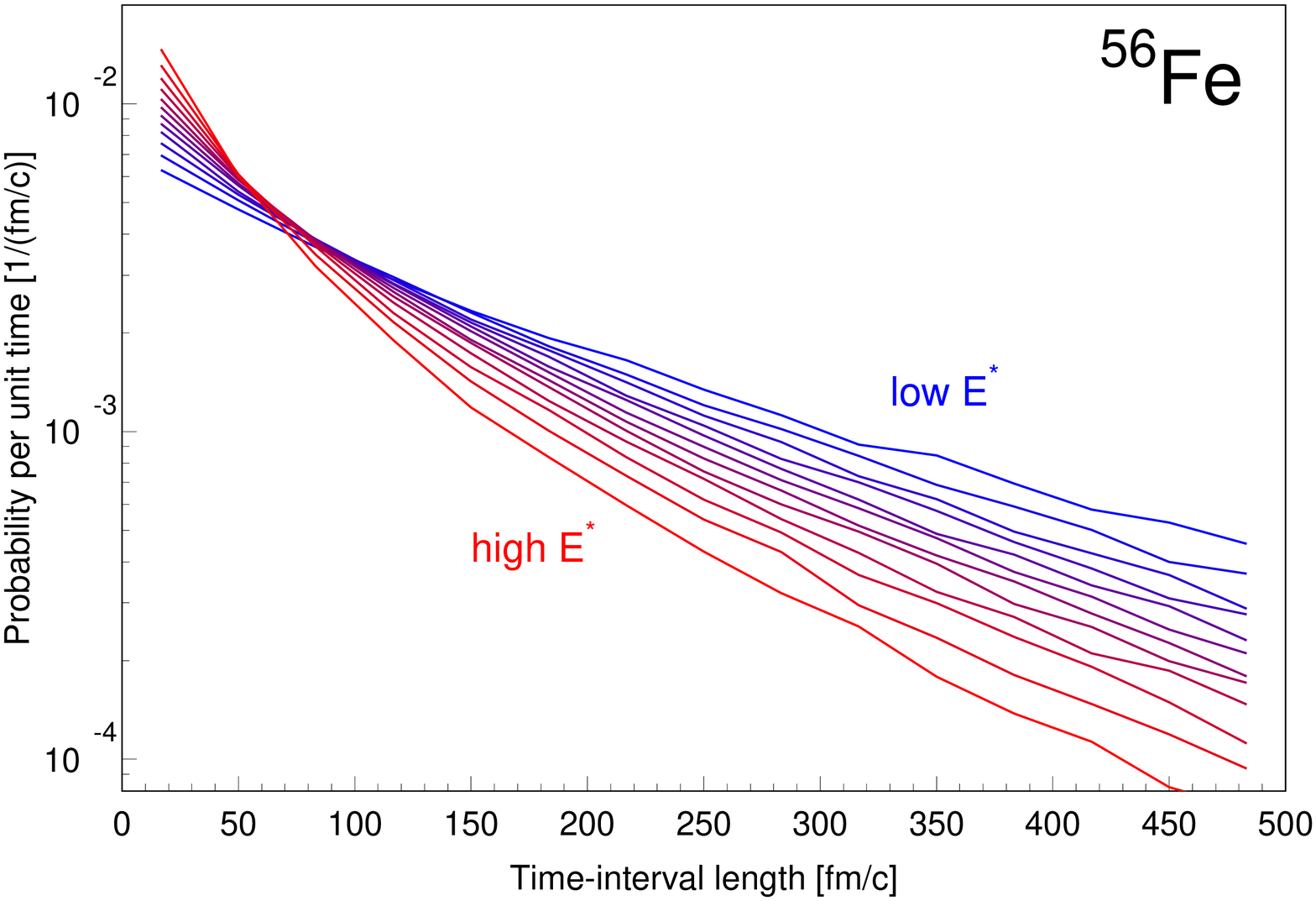}\,\,%
  \includegraphics[height=0.34\linewidth]{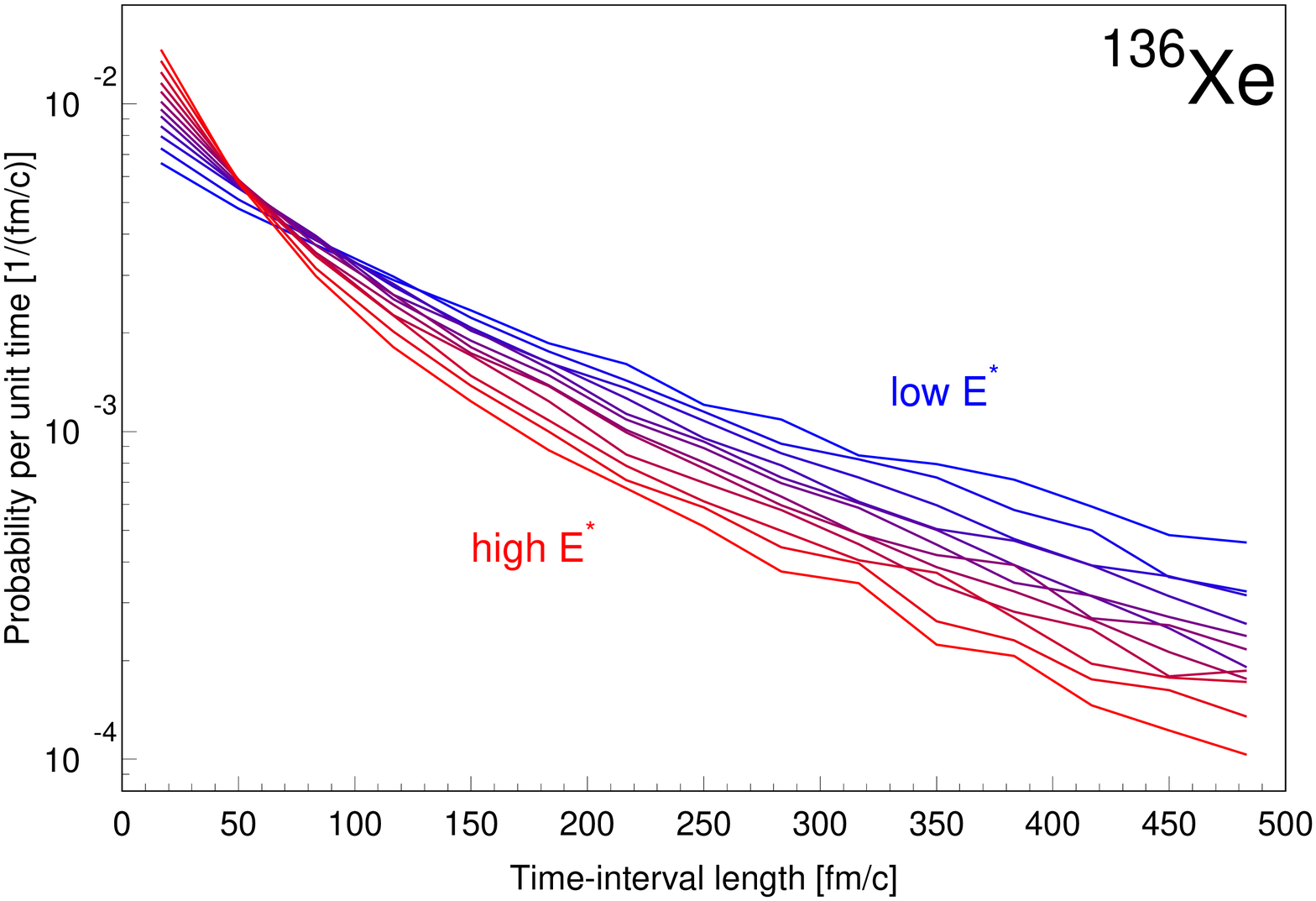}
  \caption{Distributions of time intervals between asymmetric splits. Curve colours, from blue to red, correspond to
    increasing values of the excitation energy. Each curve is normalised to one. Left panel: \proton+\iron. Right panel:
    \proton+\xenon.}
  \label{fig:deltat_distributions}
\end{figure}

Fig.~\ref{fig:deltat_distributions} displays the distributions of time intervals for the twelve excitation-energy
bins. The lowest excitation-energy bins are given above. The highest excitation-energy bins range from 8.8~$A$MeV (\iron)
and 4.6~$A$MeV (\xenon) to infinity. One notices that the distributions corresponding to low excitation energies are close
to exponential. All distributions are very broad, with heavy tails extending up to several thousands fm/$c$; for this
reason, it is inappropriate to characterise the distributions using their mean and/or their standard deviation. In what
follows, we will rather rely on the median and the interquartile range, which are more robust.

\begin{figure}
  \centering
  \includegraphics[width=0.7\linewidth]{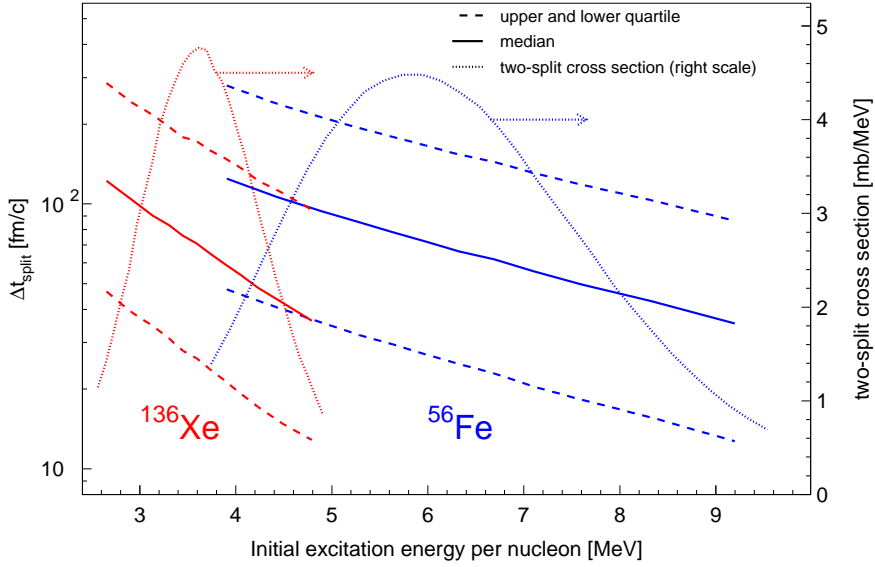}
  \caption{Median (solid lines) and quartiles (dashed lines) of the distributions of time-interval lengths between fragment
    emissions ($\Delta t_\text{split}$), as a function of the initial excitation energy per nucleon. The model is
    \incl/\geminipp. Red (blue) lines correspond to the \proton+\xenon\ (\proton+\iron) reaction. Cross sections for
    two-split events are superimposed as dotted lines (right scale).}
  \label{fig:time}
\end{figure}

The dotted lines in Fig.~\ref{fig:time} represent the energy-differential cross section for two-split events (right
scale). One immediately notices that two-split events are concentrated at higher excitation energies per nucleon in \iron\
than in \xenon. The \iron\ distribution extends up to very high excitation energies, comparable to or larger than the
total binding energy of the remnant. Such remnants are rare but nevertheless possible. Interestingly, the maxima of
the two-split cross sections are located at about 5.8 (\iron) and 3.6~$A$MeV (\xenon), which are similar to \abla's
multifragmentation thresholds cited in Sec.~\ref{sec:production-mechanism}. This strengthens the (perhaps coincidental)
similarity between \abla's multifragmentation and \geminipp's two-split events that was observed in
Sec.~\ref{sec:production-mechanism}.

On the left scale of Fig.~\ref{fig:time} we report the medians (solid lines) and quartiles (dashed) of the interval-length
distributions in each excitation-energy bin. The fact that the quartile curves are approximately parallel to the median
curve indicates that, up to a scale factor (note the logarithmic scale on the $\Delta t_\text{split}$ axis), the shapes of
the distributions are approximately the same in all the bins. If we take the median as a measure of location, we can
observe that typical interval lengths decrease as the initial excitation energy increases. Note also that, for the same
excitation energy per nucleon, interval lengths are larger for \iron\ than for
\xenon.

\incl/\geminipp\ thus predicts that typical intervals between fragment emissions last a few hundreds fm/$c$ at the onset
and reach 70--75~fm/$c$ at the peak of the two-split cross section (independently of the target). For higher excitation
energies, even shorter times are expected. These numbers are comparable to the typical multifragmentation timescale of a
few tens of fm/$c$ \cite{viola-isis,borderie-multifragmentation_review}, suggesting a continuous transition between
sequential binary decay at low energy and the expected multifragmentation regime at high energy. This aspect suggests that
a binary-decay model can generate final states that similar to those that a multifragmentation model would produce.

The time interval between fragment emissions represents an upper bound for the interval between any two consecutive binary
decays. As the excitation energy increases, this time eventually becomes comparable to the relaxation time of the
system. Under these conditions it is difficult to justify the compound-nucleus hypothesis, which assumes a completely
equilibrated system. However, it is difficult to provide quantitatively accurate estimates of the relaxation time of a
highly excited nuclear system. Besides equilibration, however, the asymptotic (observable) escape velocities of the
emitted charged particles are sensitive to the length of the interval between emissions \cite{viola-isis} and should in
principle be determined from the solution of the equations of motion of the emitted fragments in their mutual Coulomb
field. In \geminipp, as in most statistical de-excitation codes, it is assumed that decay products have already attained
their asymptotic velocity before they undergo any subsequent decay. The importance of an exact solution could then be
evaluated by studying observables that are sensitive to the de-excitation kinematics, such as LVDs. Note however that
\smm\ does include a numerical solver for the Coulomb trajectories of the hot fragments, but it is still unable to
reproduce the experimental LVDs for \proton+\xenon.

\section{Conclusions}\label{sec:conclusions}

We have used the tools of coupled intranuclear-cascade and statistical-de-excitation models to search the 1-GeV
\proton+\iron\ and \proton+\xenon\ reactions for signatures of multifragmentation. The choice of the cascade model has
some influence on the distributions of remnant size and excitation energy; in particular, dynamical emission of clusters
during the cascade stage has a sensible influence on the remnant-mass distribution and, thus, on the residue-production
cross sections. This leads to a rather large sensitivity of calculated residue-production cross sections on the cascade
model. For the purpose of this study, we chose to fix the cascade model by requiring that it correctly reproduce
residue-production cross sections close to the target nuclide, which are typically understood as due to the evaporation of
lowly-excited cascade remnants and are rather insensitive to the choice of the de-excitation model.

Calculations indicate that the inclusion of a multifragmentation stage is not crucial for adequate prediction of
residue-production cross sections. We thus confirm the insensitiveness of this observable to the de-excitation
mechanism. However, different de-excitation models propose widely different reconstructions of the residue-production
cross sections in terms of elementary processes, suggesting that semi-exclusive observables can help discriminate among
different de-excitation mechanisms. Comparisons with measured fragment-helium correlations, $Z_1-Z_2$ distributions and
IMF-gated $Z_\text{bound}$ distributions \cite{legentil-fe} favour binary de-excitation models such as \geminipp, or
models predicting very small multifragmentation cross sections such as \abla. We conclude that a multifragmentation model
is not necessary for the description of inclusive and semi-exclusive observables. This does not mean that the presence of
multifragmentation is ruled out in these reactions, but rather that binary decay can generate final states similar to
those produced by multifragmentation models, at least close to the multifragmentation threshold.

Somewhat ambiguous conclusions can be drawn from the qualitative study of longitudinal-velocity distributions. Contrary to
previous claims \cite{napolitani-fe}, we find that pure binary decay can account for the distributions measured in
\proton+\iron. The observed single-peaked component can be ascribed to multifragmentation, de-excitation residues, or
both, depending on the de-excitation model considered. On the other hand, \emph{none} of the considered de-excitation
models can explain the existence of the observed single-peaked component in \proton+\xenon. Even numerical integration of
the Coulomb trajectories of the multifragmentation products, as implemented in \smm,
predicts double-peaked longitudinal-velocity distributions. Therefore, the shape of the longitudinal-velocity distribution
is an ambiguous signature of de-excitation mechanism.

The \incl/\geminipp\ calculations suggest that \proton+\xenon\
residues with $4\leq Z\lesssim40$ are mostly produced in events with one asymmetric split,
with no contribution from de-excitation residues; it is far from obvious that a
similar mechanism can produce single-peaked longitudinal-velocity distributions. If the single-peaked component in
\proton+\xenon\ must be ascribed to multifragmentation, we would expect the multifragmentation signature to be even more
visible in \proton+\iron; but that does not appear to be the case.

We have also studied the time interval between asymmetric splits in \incl/\geminipp. At the highest excitation energies
per nucleon, the model predicts interval lengths comparable with the typical multifragmentation timescale. This again
suggests a smooth transition between the binary-decay and the multifragmentation regimes and illustrates how binary decay
can generate multifragmentation-like final states, as mentioned above. It is not clear whether equilibration times of the
order of the time interval between asymmetric splits are sufficiently long to justify \geminipp's compound-nucleus
hypothesis; in any case, closely-packed binary emissions of charged fragments are expected to distort the asymptotic
Coulomb velocities. This effect is not accounted for in any of the models considered in the present work.

In conclusion, binary decay yields a satisfactory description of most of the observables considered in this paper. The
application of binary-decay models to cascade remnants with very large excitation energies generates final states that
resemble those produced by multifragmentation models. The good agreement of \incl/\geminipp\ and \incl/\abla\ with the
experimental data considered in this paper probably indicates that events with very high excitation energy per nucleon do
not significantly contribute to the studied observables.

\begin{acknowledgments}
  We thank the authors of the de-excitation models (A.~Botvina, R.~J.~Charity, A.~Keli\'{c}-Heil, M.~V.~Ricciardi and
  K.-H.~Schmidt) for enlightening discussions. P.~Kaitaniemi's help for the \code{Geant4} simulation of the
  SPALADIN setup is gratefully acknowledged.
\end{acknowledgments}

\end{document}